\documentclass[final,5p,times,twocolumn]{elsarticle}

\makeatletter
\def\ps@pprintTitle{
   \let\@oddhead\@empty
   \let\@evenhead\@empty
   \def\@oddfoot{}
   \let\@evenfoot\@oddfoot}
\makeatother

\usepackage{amsmath}
\usepackage{braket}
\usepackage{mathtools}
\usepackage{tikz}
\usepackage{xcolor}
\usepackage{tabularx}
\usepackage{array}
\usepackage{makecell}
\newcolumntype{C}{>{\centering\arraybackslash}m{3.0cm}}
\usepackage{xcolor}
\usepackage{float}\floatstyle{ruled}\newfloat{algorithm}{tbp}{loa}\floatname{algorithm}{Algorithm}
\usepackage{booktabs}
\usepackage[T1]{fontenc}
\usepackage[utf8]{inputenc}
\usepackage{graphicx,microtype}
\usepackage{placeins,stfloats}
\usepackage[hidelinks]{hyperref}
\biboptions{numbers,sort&compress}
\usepackage[T1]{fontenc}
\usepackage[utf8]{inputenc}
\usepackage{amsmath,amssymb,mathtools,booktabs,graphicx,microtype}
\usepackage{xcolor,tikz,stfloats,placeins}
\usepackage{algorithm}
\usepackage{algpseudocode}
\usetikzlibrary{arrows.meta}

\definecolor{BBbranch}{HTML}{245A81}
\definecolor{BBprune}{HTML}{A33832}
\definecolor{BBupdate}{HTML}{17694D}

\usepackage{xspace,etoolbox}

\makeatletter
\ifdefempty{\elsauthors}{
 \patchcmd{\MaketitleBox}{\vskip36pt}{\vskip0pt}{}{}
}{}
\makeatother

\newcommand{\Sph}{\mathbb{S}}

\newcommand{\Local}{L_1}

\newcommand{\KG}{K_{\mathrm G}}
\newcommand{\normone}[1]{\left\lVert #1\right\rVert_1}
\newcommand{\pmone}{\{\pm1\}}

\newcommand{\BB}{branch-and-bound\xspace}
\newcommand{\Lnorm}[1][]{
  \if\relax\detokenize{#1}\relax
    L_1
  \else
    L_1(#1)
  \fi
}

\usepackage{amssymb}
\newcounter{bla}

\allowdisplaybreaks
\begin{document}

\begin{frontmatter}

\title{Exact classical bounds of bipartite correlation Bell inequalities using improved branch-and-bound}

\author[a]{Istv\'an M\'arton\corref{author}}
\author[b]{P\'eter Divi\'anszky}
\author[a]{Erika Bene}
\author[a]{Tam\'as V\'ertesi}

\cortext[author] {Corresponding author.\\\textit{E-mail address:} marton.istvan@atomki.hu}
\address[a]{HUN-REN Institute for Nuclear Research, P.O. Box 51, Debrecen, H-4001, Hungary}
\address[b]{Faulhorn Labs, Budafoki út 91-93, Budapest, H-4117, Hungary}

\begin{abstract}
We present an improved branch-and-bound (IBB) algorithm for the exact
computation of classical (local) bounds of bipartite correlation Bell inequalities.
We strengthen pruning in the standard BB algorithm in two ways: a heuristic search based on see-saw iteration supplies a strong initial lower bound, and greedy row reordering of the Bell matrix reduces the tail
bounds used during the exact search. An OpenMP implementation combines
these improvements with shared-memory parallelism and dynamic work
redistribution. Random-matrix benchmarks show substantial runtime reductions in matrix sizes, while explicit constructions demonstrate exact local-bound computations for square matrices up to $130\times130$ and
rectangular matrices of sizes up to $81\times 321$. Applications provide improved
lower bounds on finite-order Grothendieck constants for $d=3,4,5,7$ and
uniform-input XOR games with quantum-to-classical ratios exceeding
$\sqrt2$. The method retains an exponential worst-case search for this NP-hard optimization problem, and its practical performance depends on matrix structure and tail-bound preprocessing depth.
\end{abstract}
\end{frontmatter}

\section{Introduction}
\label{sec:intro}

Quantum theory predicts correlations between spatially
separated systems that cannot be reproduced by local classical models. This phenomenon, known as Bell nonlocality, is detected through violations of
Bell inequalities, which place limits on the correlations
allowed by such local classical models~\cite{Bell1964,Brunner2014}.

Beyond its foundational significance, Bell nonlocality is
an important resource for quantum information processing,
particularly in device-independent protocols, where conclusions
are drawn from observed measurement statistics without relying
on a detailed model of the devices~\cite{Scarani2012}.
Applications include secure key
distribution~\cite{AcinPRL2007}, reducing communication
requirements in distributed computational
tasks~\cite{BuhrmanRMP2010}, and self-testing and certification
of quantum states and
measurements~\cite{SupicQuantum2020,YangPRL2014}.
Beyond these quantum information tasks, nonlocal correlations
can also provide advantages in Bayesian
games~\cite{BrunnerLinden2013,KoniorczykPRA2020}
and in coordination tasks, including rendezvous on
networks~\cite{MironowiczNJP2023} and load balancing
in networked systems~\cite{ArunHotNets2025}, among many others.

An important ingredient in the analysis of these
applications is determining the limits of local classical
strategies. Here we focus on bipartite Bell expressions
with two outcomes per measurement and containing only
correlations between the parties' outcomes. In the language
of theoretical computer science, these expressions correspond
to XOR games~\cite{Cleve2004}. Their coefficients form
a Bell matrix $M$. We study the exact computation of its
classical bound, with particular emphasis on matrices
corresponding to many measurement settings, for which exact optimization can be computationally
demanding. We expect that
the algorithmic approach can also be adapted to other Bell
scenarios, including measurements with more than two outcomes,
by developing suitable branching rules and pruning bounds.

For a real Bell matrix $M\in\mathbb{R}^{n\times m}$, the classical (or local) bound of the associated two-outcome bipartite full-correlation Bell expression is~\cite{Bell1964,Brunner2014}

\begin{equation}
\label{CB}
    \Lnorm \left( M \right) = \max_{a\in\pmone^n,\,b\in\pmone^m} \sum_{y=1}^m \sum_{x=1}^n M_{xy} a_x b_y .
\end{equation}
Throughout this paper, we refer to this bound interchangeably as the local or classical value; the latter terminology is more natural in the computer science context.

Optimizing Bob's signs for fixed $a$ gives~\cite{DivianszkyPRA2017,MartonCPC2025}
\begin{equation}
\label{eq:l1man}
  \begin{aligned}
    \Lnorm[M]&=\max_{a\in\pmone^n}
       \sum_{y=1}^m\left|\sum_{x=1}^n M_{xy}a_x\right|\\
    &=\max_{a\in\pmone^n}\normone{M^{\mathsf T}a}.
  \end{aligned}
\end{equation}
where $\normone{v}=\sum_i|v_i|$ is the Manhattan norm.

After transposition, if necessary, we may assume $n\le m$. Global sign symmetry allows one sign, say $a_1$, to be fixed to $+1$, reducing exhaustive enumeration to $2^{n-1}$ assignments. Gray-code updates then yield an $O(m2^{n-1})$ exact algorithm rather than the naive enumeration of all $2^{n+m}$ pairs of strategies~\cite{MartonCPC2025}. Nevertheless, the exponential dependence remains prohibitive for large $n$. More generally, the NP-hardness of the problem precludes a polynomial-time exact algorithm for arbitrary instances unless $P=NP$.

The computational complexity of the binary bilinear optimization in Eq.~\eqref{eq:l1man} is studied in Refs.~\cite{AlonSIAM2006,KhotCoPaAM2012}. A related hardness result for classical correlations is the NP-completeness of correlation-polytope membership~\cite{Pitowsky1991}. Exact local bounds are particularly valuable in searches for improved lower bounds on finite-order Grothendieck constants: combined with explicit vector strategies, they yield rigorous quantum-to-classical ratios.

The real Grothendieck constant~\cite{GrothendieckBSMSP1953} bounds the advantage of vector correlations
over binary strategies in bipartite full-correlation expressions.
Its finite-order counterpart restricts the real vector dimension to $d$,
and $K_G=\lim_{d\to\infty}K_G(d)$.
The vector-valued, or Tsirelson~\cite{TsirelsonJoSM1987}, optimization is
\begin{equation}
Q(M,d)=\max_{\substack{u_x,v_y\in\Sph^{d-1}}}
  \sum_{x=1}^{n}\sum_{y=1}^{m}M_{xy}\, \langle u_x, v_y\rangle.
  \label{eq:quantum-vector-bound}
\end{equation}
The real Grothendieck constant of order $d$ is~\cite{AcinPRA2006}
\begin{equation}
  \KG(d)=\sup_{n,m\ge1}\ \sup_{M\ne0}\frac{Q(M,d)}{\Local(M)}.
  \label{eq:kgd}
\end{equation}
Here $d$ is the real vector dimension in Tsirelson's representation, not the local Hilbert-space dimension. A feasible set of vectors gives a value $q_d(M)\le Q(M,d)$ and therefore the rigorous lower bound
\begin{equation}
  \KG(d)\ge \frac{q_d(M)}{\Local(M)}.
  \label{eq:kg-lower}
\end{equation}
This connection links Grothendieck's inequality~\cite{GrothendieckBSMSP1953,LindenstraussSM1968,PisierBAMS2012,Oertel2024} to Bell nonlocality~\cite{TsirelsonJoSM1987,AcinPRA2006,Brunner2014}. The exact identity $\KG(2)=\sqrt2$ is known~\cite{KrivineAdvinMat1979}, whereas the exact values for $d\ge3$, including the limit $d\to\infty$, remain open. For $d=3$, the best published rigorous bounds are $1.43670\le\KG(3)\le1.45463$~\cite{DesignollePRA202602,Designolle2023}; see also the recent discussion in Ref.~\cite{Bene2026}. A more recent, unpublished result improves the upper bound
to $\KG(3)\le1.44705$~\cite{Porto2026}.
On the other hand, recent work has substantially tightened both the lower and upper bounds on the unrestricted real Grothendieck constant~\cite{JonesArxiv2026,LiArxiv2026,HeilmanJonesMalavolta2026,PengFourier2026}.

For a fixed Bell matrix, finding vector strategies for
Eq.~\eqref{eq:quantum-vector-bound} that yield large values of
$q_d(M)$ is often relatively straightforward. The harder task
is to obtain a tight upper bound on the classical value in the
denominator of Eq.~\eqref{eq:kg-lower}, or to compute that value
exactly. Exact local-bound solvers therefore play an important
role in numerical searches for Bell inequalities and improved
lower bounds on finite-order Grothendieck
constants~\cite{VertesiPRA2008,Designolle2023,DesignollePRA202602}.

Earlier numerical approaches include linear-programming methods
for testing local-model feasibility and determining
nonclassicality thresholds~\cite{Gruca2010,Gondzio2014}.
The convex-separation method of Brierley
et~al.~\cite{Brierley2017} combines heuristic searches for Bell
inequalities with exact evaluation of $\Lnorm[M]$ to verify
their classical bounds. Their reported $42\times42$ example
required approximately one week of exhaustive enumeration,
illustrating the substantial computational cost even for matrices considerably smaller than the largest instances considered here.

The present work builds on the exact algorithms of
Refs.~\cite{MartonCPC2025,DivianszkyPRA2017} to extend the
range of Bell matrices whose classical bounds can be computed
exactly. A branch-and-bound (BB) search~\cite{LandDoig1960}
explores deterministic strategies and discards an entire branch
whenever an upper bound shows that it cannot improve on the
best classical value already found. Our improved BB algorithm
(IBB) makes this pruning more effective in two ways. First, a see-saw search finds a good deterministic strategy before the exact search begins, providing a strong initial lower bound on $L_1(M)$. Second, a greedy reordering of the matrix rows aims to tighten the upper bounds on the contributions from the remaining rows. These improvements can substantially reduce search effort without changing the binary optimum. Their benefit depends on the instance and on the cost of preprocessing, however, they do not remove the exponential worst-case complexity. The implementation is written in C and uses OpenMP~\cite{OpenMP} for shared-memory parallelism. We demonstrate its performance on random
matrices and on explicit Bell matrices, including a $130\times130$
signed instance and rectangular instances of sizes $81\times321$ and $72\times360$. The exact local-bound computations yield improved lower bounds on finite-order Grothendieck constants $K_G(d)$ for $d=3,4,5,7$, including $K_G(3)\ge 1.438746$. We also construct five explicit signed Bell matrices, with entries restricted to ${\pm1}$, whose associated uniform-input XOR games achieve quantum-to-classical bias ratios exceeding the CHSH value~$\sqrt{2}$.

The remainder of the paper is organized as follows.
Section~\ref{sec:BB} reviews the exact BB search and its memory
requirements. Section~\ref{sec:IBB} introduces see-saw initialization
and greedy row reordering in Secs.~\ref{sec:IBB-seesaw}
and~\ref{sec:IBB-reordering}, evaluates single-thread performance in
Sec.~\ref{sec:singleIBB}, and describes the parallel implementation in
Sec.~\ref{sec:paraIBB}. Section~\ref{sec:results} presents applications
to finite-order Grothendieck bounds and signed Bell games in
Secs.~\ref{sec:applications-grothendieck} and~\ref{app:signed_games},
respectively. Section~\ref{sec:disc} summarizes the results and discusses remaining computational challenges and open questions in quantum foundations.

\FloatBarrier
\section{The \BB algorithm} \label{sec:BB}

\subsection{Principles of the branch-and-bound algorithm}
\label{sec:BB-recap}

Branch-and-bound (BB)~\cite{LandDoig1960} organizes a discrete
optimization problem as a search tree and discards subtrees that cannot
improve the current lower bound on the maximum. In this section, we describe the implementation of Divi\'anszky et~al.~\cite{DivianszkyPRA2017} for the
local bound in Eq.~\eqref{CB}. Writing $M_x$ for row $x$ of an
$n\times m$ Bell matrix, optimizing Bob's signs reduces the objective
to the Manhattan norm of the signed row sum
$\sum_{x=1}^n a_xM_x$. Here $\Lnorm[M]$ denotes the local bound of
the matrix, whereas $\|v\|_1=\sum_{y=1}^m|v_y|$ denotes the
Manhattan norm of a vector, consistently with
Ref.~\cite{MartonCPC2025}. Reversing all of Alice's signs leaves
this norm unchanged, so we fix $a_1=+1$ and search over the remaining
$n-1$ signs.

\paragraph{Search states and the recursion}
A node at depth $k$ represents a partial strategy
$(a_1,\ldots,a_k)$ and stores the partial sum
$v=\sum_{x=1}^k a_xM_x$. Its two children assign the next sign:
$a_{k+1}=+1$ gives $v+M_{k+1}$, while $a_{k+1}=-1$ gives
$v-M_{k+1}$. The search also maintains an \emph{incumbent} $B$,
a lower bound on the global maximum. This can be initialized from
a known strategy or another valid lower bound and is increased
whenever a better strategy is found.

To decide whether to explore a node, choose tail coefficients
\[
 c_k\geq\Lnorm[M_{k+1:n}]\quad(1\leq k<n),
 \qquad c_n=0,
\]
where $M_{k+1:n}$ contains rows $k+1$ through $n$. As justified
below, $U_k(v)=\|v\|_1+c_k$ bounds the best value attainable by
completing the current partial strategy. The depth-first search is
then expressed by
\begin{equation}
 \label{BB_condition}
 \begin{aligned}
  &g_{M,c}(k,v,B)={}\\[-.5ex]
  &\;\begin{cases}
   B, & U_k(v)\leq B,\\
   \|v\|_1, & k=n\text{ and }\|v\|_1>B,\\
   g_{M,c}(k+1,v-M_{k+1},B_+), & \text{otherwise},
  \end{cases}
 \end{aligned}
\end{equation}
where, in the branching case,
\[
 B_+=g_{M,c}(k+1,v+M_{k+1},B).
\]
The arguments $(k,v,B)$ specify the depth, partial sum, and
incumbent on entry. The function returns the larger of the incoming
incumbent and the best value attainable within that subtree.
Its three cases have the following interpretation:
\begin{enumerate}
 \item \emph{Prune.} If $U_k(v)\leq B$, no completion can improve
 the incumbent. Return $B$ without visiting any descendants.
 \item \emph{Update at a leaf.} If all signs have been assigned
 ($k=n$), then $c_n=0$ and $\|v\|_1$ is the value of a complete
 strategy. If it exceeds $B$, return this improved value.

\item\emph{Branch.}
If the node has not been pruned and $k<n$, create two child
nodes at depth $k+1$, corresponding to the choices
$a_{k+1}=-1$ and $a_{k+1}=+1$. Their partial-sum vectors are
$v-M_{k+1}$ and $v+M_{k+1}$, respectively. Replace the current
array entry with the $-1$ child and store the $+1$ child in
the next position, advancing the index of the last active
entry. Since the algorithm processes the last active entry,
the $+1$ subtree is explored first.
Branching itself does not change the global lower bound $B$.
This value is updated only when a leaf with a larger objective
value is reached. When the search subsequently explores the
$-1$ subtree, it uses the current value of $B$, including any
improvement found in the $+1$ subtree. In the recursive notation
of Eq.~\eqref{BB_condition}, $B_+$ denotes the best value
available after processing the $+1$ subtree; in the
implementation, this value is maintained in the global
variable storing $B$. The corresponding array operations
are illustrated in Table~\ref{table_memory}.
\end{enumerate}
Thus an improvement found in one branch can enable pruning in a
branch visited later. Equality is sufficient for pruning when only
the optimum value is required, even if a discarded subtree contains
another strategy attaining that value.

Starting at the first row gives
$\Lnorm[M]=g_{M,c}(1,M_1,B_0)$ for any valid initial bound
$0\leq B_0\leq\Lnorm[M]$; $B_0=0$ is always permissible.
Every discarded subtree has an upper bound no larger than the
incumbent, and every remaining leaf is evaluated. The final result
is therefore the global optimum. The worst-case search remains
exponential; pruning reduces the work only when the bounds are
sufficiently restrictive.

\paragraph{Why the pruning bound is valid}
The row-partition inequality of Ref.~\cite{DivianszkyPRA2017} states
\begin{equation}
 \label{inequality}
 \Lnorm[M]\leq\Lnorm[M^U]+\Lnorm[M^L],
 \qquad M=\begin{pmatrix}M^U\\M^L\end{pmatrix}.
\end{equation}
At depth $k$, let $M^U$ contain the first $k$ rows and $M^L$ the
remaining rows. The signs in the upper block are already fixed,
so its contribution is the particular vector $v$. For any
completion, the triangle inequality gives
\begin{equation*}
 \begin{aligned}
 \left\lVert v+\sum_{x=k+1}^n a_xM_x\right\rVert_1
 &\leq \|v\|_1+
       \left\lVert\sum_{x=k+1}^n a_xM_x\right\rVert_1\\
 &\leq \|v\|_1+\Lnorm[M_{k+1:n}]\\
 &\leq \|v\|_1+c_k=U_k(v).
 \end{aligned}
\end{equation*}
This is the node-specific version of Eq.~\eqref{inequality}:
we use the actual partial sum for the assigned rows and bound
only the contribution of the unassigned rows.

The smallest admissible coefficients in this bound are the exact
tail optima $c_k=\Lnorm[M_{k+1:n}]$. They can be computed by BB
from shorter tails to longer ones, starting with $c_n=0$ and using
the previously computed coefficients in each subsequent search.
To limit preprocessing, exact values may instead be computed only
for the shorter tails, with looser upper bounds elsewhere.
For example, $c_k=\sum_{x=k+1}^n\|M_x\|_1$ is always admissible,
whereas $c_k=+\infty$ disables pruning at that depth. Note that a heuristic
\emph{lower} estimate of a tail optimum cannot replace an upper
bound: doing so could discard a subtree containing a better
strategy.

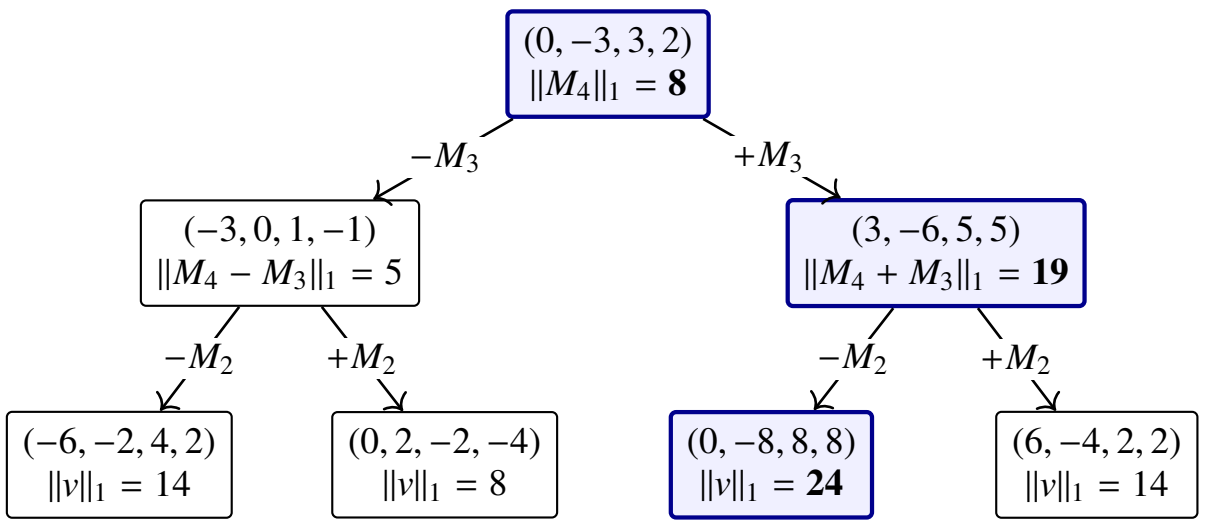
\begin{figure}[tbp]
  \centering
  \begin{tikzpicture}[
    x=\columnwidth,y=1cm,
    every node/.style={font=\scriptsize,align=center},
    box/.style={draw,rounded corners=1pt,inner sep=3pt},
    best/.style={box,draw=blue!55!black,fill=blue!6,line width=.8pt},
    edge/.style={->,line width=.5pt},
    lab/.style={fill=white,inner sep=1pt}
  ]
    \node[best] (r) at (0,0)
      {$(0,-3,3,2)$\\$\|M_4\|_1=\mathbf{8}$};
    \node[box] (l) at (-.245,-1.25)
      {$(-3,0,1,-1)$\\$\|M_4-M_3\|_1=5$};
    \node[best] (h) at (.245,-1.25)
      {$(3,-6,5,5)$\\$\|M_4+M_3\|_1=\mathbf{19}$};
    \node[box] (ll) at (-.365,-2.65)
      {$(-6,-2,4,2)$\\$\|v\|_1=14$};
    \node[box] (lr) at (-.122,-2.65)
      {$(0,2,-2,-4)$\\$\|v\|_1=8$};
    \node[best] (hl) at (.122,-2.65)
      {$(0,-8,8,8)$\\$\|v\|_1=\mathbf{24}$};
    \node[box] (hr) at (.365,-2.65)
      {$(6,-4,2,2)$\\$\|v\|_1=14$};
    \draw[edge] (r) -- node[lab,pos=.48] {$-M_3$} (l);
    \draw[edge] (r) -- node[lab,pos=.48] {$+M_3$} (h);
    \draw[edge] (l) -- node[lab,pos=.50] {$-M_2$} (ll);
    \draw[edge] (l) -- node[lab,pos=.50] {$+M_2$} (lr);
    \draw[edge] (h) -- node[lab,pos=.50] {$-M_2$} (hl);
    \draw[edge] (h) -- node[lab,pos=.50] {$+M_2$} (hr);
  \end{tikzpicture}
  \caption{Exact tail bounds for the matrix in
  Eq.~\eqref{example_matrix}. Each node displays a signed row sum
  and its Manhattan norm; an edge adds or subtracts the indicated
  row. The largest value at each level is shown in bold in a shaded
  box, giving $c_3=8$, $c_2=19$, and $c_1=24$. This small tree is
  fully enumerated for illustration; larger tail problems are
  themselves solved by branch-and-bound. Row indices start at one.}
  \label{fig:bnb-diagram}
\end{figure}

\paragraph{An explicit example}
Consider the $4\times 4$ matrix used in Ref.~\cite{DivianszkyPRA2017}:
\begin{equation}
 \label{example_matrix}
 M=\begin{pmatrix*}[r]
  2 &  3 &  3 &  0\\
  3 &  2 & -3 & -3\\
  3 & -3 &  2 &  3\\
  0 & -3 &  3 &  2
 \end{pmatrix*}.
\end{equation}
Figure~\ref{fig:bnb-diagram} enumerates the signed sums of its last
one, two, and three rows. For each of these tail problems, reversing all row signs changes
the signed sum from $v$ to $-v$ and leaves its Manhattan norm
unchanged. We may therefore fix any one row sign and choose
the sign of $M_4$ to be $+1$. This uses the same symmetry as
fixing $a_1=+1$ in the full search, with a different row chosen because the first row is absent from these tails. Taking the largest Manhattan norm at each level yields
\begin{equation}
 \label{matrix_with_coefficients}
 \begin{aligned}
 c_1&=\Lnorm[M_{2:4}]=24, & c_2&=\Lnorm[M_{3:4}]=19,\\
 c_3&=\|M_4\|_1=8,       & c_4&=0.
 \end{aligned}
\end{equation}
In particular, $c_3$ bounds only the remaining fourth row once
three signs have been assigned.

For this example, $B_0=c_1=24$ is a valid initial lower bound.
Indeed, for any signed sum $t$ of the last three rows, the triangle
inequality implies
\[
 \max\{\|t+M_1\|_1,\|t-M_1\|_1\}\geq\|t\|_1.
\]
Maximizing over $t$ proves that adding the first row cannot reduce
the local bound. Here we use the \emph{exact} tail value $c_1$;
an arbitrary upper bound on that tail would not justify this
initialization.

The prefix $(a_1,a_2,a_3)=(1,1,1)$ gives $v=(8,2,2,0)$ and
$U_3(v)=12+8=20\leq24$, so both choices of $a_4$ can be skipped.
The complete strategy $(a_1,a_2,a_3,a_4)=(1,1,-1,-1)$ instead gives $v=(2,11,-5,-8)$ and
$\|v\|_1=26$, increasing the incumbent to 26. This improvement
also permits pruning the prefix $(1,-1,1)$, whose bound is
$U_3(v)=26$: it could not be pruned with the initial incumbent 24.
Completing the search proves $\Lnorm[M]=26$.

A step-by-step account of the full traversal, including all pruning
decisions and incumbent updates, is given in~\ref{BB_algorithm_with_example}.

\subsection{Memory handling}
\label{sec:memory}

The depth-first recursion in Eq.~\eqref{BB_condition} can be
implemented iteratively using a last-in, first-out stack of pending
nodes. The active node is the last occupied entry. When a node
branches, its record is replaced by the $-1$ child and the $+1$
child is appended, so that the latter is processed first. A pruned
node is removed immediately. At a leaf, the current lower bound
$B$ is updated if necessary, and the node is then removed. This
ordering reproduces the recursive evaluation of $g_{M,c}$ without
storing the entire search tree.

The required capacity is determined by the search depth. If each
branching step replaces one node by $q$ children, it increases the
number of pending nodes by $q-1$. During depth-first traversal,
at most $q-1$ unexplored siblings remain at each ancestral level.
Starting at $k=1$, there are at most $n-1$ such levels, giving the
worst-case capacity
\begin{equation}
  \label{eq:memory-capacity}
  N_{\mathrm{cap}}=1+(n-1)(q-1)
  \underset{q=2}{=}n.
\end{equation}
Thus the binary search requires space for at most $n$ pending
nodes, although pruning can reduce the peak occupancy. Here $q$
denotes the branching factor, to distinguish it from the strategy
variables $a_x$.

In this representation, a node record contains the depth $k$, the
partial sum $v\in\mathbb{R}^{m}$, and the assigned signs
$S=(a_1,\ldots,a_k)$. A fixed-capacity record reserves $m$ entries
for $v$ and $n$ entries for $S$, of which only the first $k$ are
used. The bound $B$ and the coefficients $c_k$ are shared by all
records. Storing up to $n$ such records requires
$O(nm+n^2)$ scalar entries, in addition to the $O(nm)$ input
matrix and the $O(n)$ coefficient array. With the preprocessing
convention $n\leq m$, the total storage is therefore $O(nm)$,
consistent with the matrix-size memory scaling discussed in
Ref.~\cite{DivianszkyPRA2017}.

In the implementation, node records are stored in a fixed-size
contiguous array whose dimensions are specified at compile time.
The same storage is reused throughout the search, so individual
nodes do not require separate memory allocation or deallocation.
The variable \texttt{counter}, denoted by $p$ in
Table~\ref{table_memory}, is the zero-based index of the last
active record and identifies the node to be processed next.

In the single-thread implementation, pending nodes are stored in a preallocated array whose capacity is fixed at compile time. The index \texttt{counter} identifies the last active entry. Pruning removes a node from consideration simply by decrementing \texttt{counter}; the stored record is neither cleared nor modified. Its contents remain in the array but are no longer part of the active search. 
Branching replaces the current entry with the $-1$ child
and writes the $+1$ child at position \texttt{counter+1},
then increments \texttt{counter} so that the $+1$ child
is processed first. Previously inactive entries can therefore be overwritten as the search proceeds. The array remains allocated throughout the search.

Contiguous storage also keeps neighbouring records close together
in memory. Since processors transfer data to their caches in
blocks called cache lines, this layout can improve cache locality
compared with separately allocated linked-list nodes. It also
avoids the additional pointers needed to link such nodes.

In the parallel implementation, each thread maintains its own
array and two indices: \texttt{counter\_back} and
\texttt{counter} mark the first and last active records,
respectively. The thread processes records from the latter end.
During work stealing, the record at \texttt{counter\_back} is
transferred to an idle thread, and the donor increments
\texttt{counter\_back}. Records at this end generally represent
shallower nodes, with more signs still unassigned. Transferring
such a node can therefore move a larger part of the remaining
search and reduce the number of transfers between threads.

For completeness, a finite alternative to the formal choice
$c_k=+\infty$ is the entrywise absolute sum
\[
  T=\sum_{x=1}^{n}\sum_{y=1}^{m}|M_{xy}|.
\]
Every tail bound satisfies $\Lnorm[M_{k+1:n}]\leq T$, so assigning
$c_k=T$ for $k<n$ is always admissible. For the example matrix,
$T=38$, and this loose choice is
\begin{equation}
  \label{matrix_with_coefficients_infinite}
  \begin{gathered}
    M=\begin{pmatrix*}[r]
      2 &  3 &  3 &  0\\
      3 &  2 & -3 & -3\\
      3 & -3 &  2 &  3\\
      0 & -3 &  3 &  2
    \end{pmatrix*},\\[.4ex]
    c=(c_1,c_2,c_3,c_4)=(38,38,38,0).
  \end{gathered}
\end{equation}
In this example, $B\leq\Lnorm[M]=26<38$, so no internal node
can satisfy $\|v\|_1+c_k\leq B$ with these coefficients. Every
complete strategy with $a_1=+1$ is therefore reached. In general,
a finite admissible coefficient is not equivalent to disabling
the pruning test. Using $+\infty$ at every internal depth does
disable pruning there and reduces the traversal to exhaustive
enumeration. The remaining bookkeeping then offers no pruning
advantage over a dedicated enumeration method, such as the
Gray-code algorithm of Ref.~\cite{MartonCPC2025}.

Table~\ref{table_memory} illustrates the stack operations using
the tighter coefficients $c=(24,19,8,0)$ from
Eq.~\eqref{matrix_with_coefficients}, rather than the loose choice
in Eq.~\eqref{matrix_with_coefficients_infinite}. Each row is a
snapshot immediately before processing the active node. The
zero-based index $p$ identifies the last occupied entry; $p=-1$
indicates an empty stack and terminates the search. The initial
lower bound is $B=24$, as justified in the preceding subsection.
The leaf with $S=(+,+,-,-)$ raises it to $26$; subsequent pruning
exhausts the remaining nodes and establishes $\Lnorm[M]=26$.

\begin{table}[tbp]
  \centering
  \begin{tikzpicture}[
    x=\columnwidth,y=-1.02cm,
    every node/.style={font=\scriptsize},
    record/.style={draw=black!55,rounded corners=1pt,
      text width=.210\columnwidth,align=center,
      inner xsep=1.3pt,inner ysep=3pt},
    active/.style={record,draw=blue!55!black,
      fill=blue!6,line width=.8pt},
    info/.style={text width=.22\columnwidth,align=center,
      inner sep=1pt}
  ]
    \node[anchor=west] at (0,-.85) {Step};
    \node at (.38,-.85) {Pending nodes (active at right)};
    \node[info] at (.87,-.85) {State / next action};
    \draw[black!50] (0,-.53) -- (.99,-.53);
    \node[anchor=center,inner sep=0pt] at (.012,0) {0};
    \node[active] at (0.155,0) {$(2,3,3,0)$\\$S=+$\\$k=1$};
    \node[info] at (.87,0) {$p=0$\\$B=24$\\branch};
    \node[anchor=center,inner sep=0pt] at (.012,1) {1};
    \node[record] at (0.155,1) {$(-1,1,6,3)$\\$S=+-$\\$k=2$};
    \node[active] at (0.38,1) {$(5,5,0,-3)$\\$S=++$\\$k=2$};
    \node[info] at (.87,1) {$p=1$\\$B=24$\\branch};
    \node[anchor=center,inner sep=0pt] at (.012,2) {2};
    \node[record] at (0.155,2) {$(-1,1,6,3)$\\$S=+-$\\$k=2$};
    \node[record] at (0.38,2) {$(2,8,-2,-6)$\\$S=++-$\\$k=3$};
    \node[active] at (0.605,2) {$(8,2,2,0)$\\$S=+++$\\$k=3$};
    \node[info] at (.87,2) {$p=2$\\$B=24$\\prune};
    \node[anchor=center,inner sep=0pt] at (.012,3) {3};
    \node[record] at (0.155,3) {$(-1,1,6,3)$\\$S=+-$\\$k=2$};
    \node[active] at (0.38,3) {$(2,8,-2,-6)$\\$S=++-$\\$k=3$};
    \node[info] at (.87,3) {$p=1$\\$B=24$\\branch};
    \node[anchor=center,inner sep=0pt] at (.012,4) {4};
    \node[record] at (0.155,4) {$(-1,1,6,3)$\\$S=+-$\\$k=2$};
    \node[record] at (0.38,4) {$(2,11,-5,-8)$\\$S=++--$\\$k=4$};
    \node[active] at (0.605,4) {$(2,5,1,-4)$\\$S=++-+$\\$k=4$};
    \node[info] at (.87,4) {$p=2$\\$B=24$\\prune};
    \node[anchor=center,inner sep=0pt] at (.012,5) {5};
    \node[record] at (0.155,5) {$(-1,1,6,3)$\\$S=+-$\\$k=2$};
    \node[active] at (0.38,5) {$(2,11,-5,-8)$\\$S=++--$\\$k=4$};
    \node[info] at (.87,5) {$p=1$\\$B=24$\\$B\leftarrow26$};
    \node[anchor=center,inner sep=0pt] at (.012,6) {6};
    \node[active] at (0.155,6) {$(-1,1,6,3)$\\$S=+-$\\$k=2$};
    \node[info] at (.87,6) {$p=0$\\$B=26$\\branch};
    \node[anchor=center,inner sep=0pt] at (.012,7) {7};
    \node[record] at (0.155,7) {$(-4,4,4,0)$\\$S=+--$\\$k=3$};
    \node[active] at (0.38,7) {$(2,-2,8,6)$\\$S=+-+$\\$k=3$};
    \node[info] at (.87,7) {$p=1$\\$B=26$\\prune};
    \node[anchor=center,inner sep=0pt] at (.012,8) {8};
    \node[active] at (0.155,8) {$(-4,4,4,0)$\\$S=+--$\\$k=3$};
    \node[info] at (.87,8) {$p=0$\\$B=26$\\prune};
    \node[anchor=center,inner sep=0pt] at (.012,9) {9};
    \node[anchor=west] at (.05,9) {\emph{empty}};
    \node[info] at (.87,9) {$p=-1$\\$B=26$\\stop};
  \end{tikzpicture}
  \caption{Depth-first stack trace with exact tail bounds
  $c=(24,19,8,0)$ and initial bound $B=24$. Boxes contain
  $v$, the sign prefix $S$ ($\pm$ denotes $\pm1$), and depth $k$.
  The shaded box is processed next; the right column lists the
  current $p$, current $B$, and next action. The update to $26$
  occurs after step~5. Peak occupancy is three records.}
  \label{table_memory}
\end{table}

\FloatBarrier
\section{The improved BB algorithm}
\label{sec:IBB}

Before preparing the BB search, the implementation can apply the exact
matrix reduction described in Ref.~\cite[Sec.~2.1.1 and
Appendix~A]{MartonCPC2025}. Zero rows and columns are removed,
and proportional rows or columns are replaced by a single row
or column, adding or subtracting them according to the sign
of the proportionality factor. The reduced matrix $M_{\rm red}$
satisfies $\Lnorm[M_{\rm red}]=\Lnorm[M]$ and may have fewer
rows or columns. We call this step \emph{matrix reduction}.
The term \emph{search preprocessing} refers here to see-saw
initialization, greedy row reordering, and the computation of
tail bounds. Greedy row reordering permutes the rows of the matrix obtained after the optional reduction, leaving its size unchanged. Matrix reduction is controlled
separately from row reordering and tail-bound computation;
disabling it does not disable these search-preparation steps.

The improved branch-and-bound (IBB) algorithm combines the exact
search described in Sec.~\ref{sec:BB} with see-saw initialization
and greedy row reordering to strengthen pruning. At depth $k$, the vector
$v=\sum_{x=1}^{k}a_xM_x$ represents the assigned signs, $B$ is the
current lower bound on $\Lnorm[M]$, and $c_k$ is an upper bound on
the local bound of the remaining rows. The subtree is discarded
whenever
\begin{equation}
  \label{eq:IBB-pruning}
  B\geq U_k(v),\qquad U_k(v)=\|v\|_1+c_k.
\end{equation}
A larger initial $B$ makes this test more effective from the start
of the search. Smaller admissible tail coefficients can further
reduce the number of visited nodes. These observations motivate
the two improvements below: a see-saw search supplies a strong
initial lower bound (Sec.~\ref{sec:IBB-seesaw}), and greedy row
reordering targets smaller tail bounds
(Sec.~\ref{sec:IBB-reordering}). Section~\ref{sec:singleIBB}
examines their combined single-thread performance and the
trade-off between preprocessing cost and search time. 

\subsection{See-saw initialization}
\label{sec:IBB-seesaw}

\begin{figure}[!t]
  \centering
  \includegraphics[width=\columnwidth]{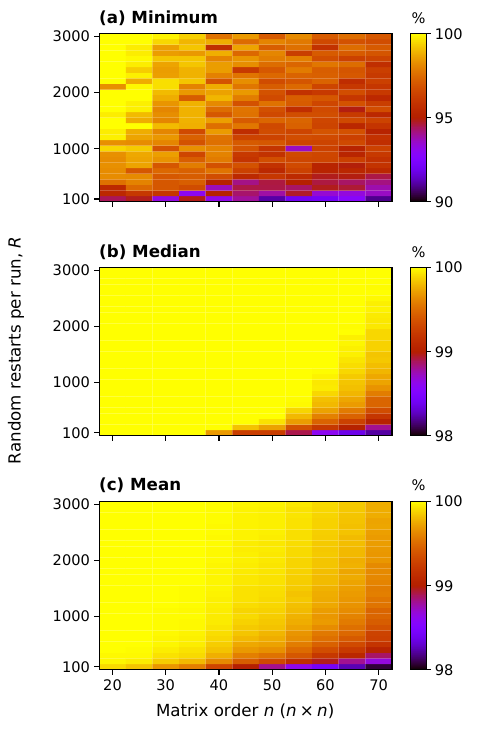}
  \caption{Quality of see-saw initialization for random
  $n\times n$ Bell matrices: (a) minimum, (b) median, and
  (c) arithmetic mean of $B_R/\Lnorm[M]$, expressed as percentages.
  Each cell summarizes 2500 runs (50 matrices, with 50 runs per
  matrix), each using $R$ random restarts. Matrix entries lie in
  $[-100,100]$. Each cell is plotted without interpolation. The color scales are 90--100\% in (a) and
  98--100\% in (b,c).}
  \label{seesaw_time}
\end{figure}

We recap the alternating optimization used as a heuristic oracle
in Appendix~A of Hirsch et~al.~\cite{HirschQuantum2017}, with the
Bell matrix $M$ as the objective coefficients. For a deterministic
strategy $\lambda=(a,b)$, the optimization problem is
\begin{equation}
  \label{see_saw}
  \begin{aligned}
    \Lnorm[M]&=\max_{\substack{a\in\{-1,+1\}^{n}\\
                              b\in\{-1,+1\}^{m}}}S_{\lambda},\\
    S_{\lambda}&:=\sum_{x=1}^{n}\sum_{y=1}^{m}M_{xy}a_xb_y.
  \end{aligned}
\end{equation}
Although the joint maximization is hard, either party's strategy
can be optimized explicitly when the other is held fixed. We
start from Bob's random strategy.
For each random restart, the procedure is as follows:
\begin{enumerate}
  \item Draw $b\in\{-1,+1\}^{m}$ uniformly at random.
  \item Optimize Alice's signs at fixed $b$, then Bob's signs at
  the updated $a$:
  \[
    \begin{aligned}
      a_x&\leftarrow\operatorname{sgn}
          \!\left(\sum_{y=1}^{m}M_{xy}b_y\right),\\
      b_y&\leftarrow\operatorname{sgn}
          \!\left(\sum_{x=1}^{n}M_{xy}a_x\right).
    \end{aligned}
  \]
  We use $\operatorname{sgn}(0)=+1$; either sign maximizes a term
  with zero coefficient.
  \item Repeat these two updates until a complete sweep gives
  no increase in $S_{\lambda}$, retaining the largest value found.
\end{enumerate}
Each update is an exact maximization over one party's signs and
therefore cannot decrease the Bell value. The stopping rule
terminates the iteration, since only finitely many deterministic
strategies exist. It does not, however, establish global
optimality. Repeating the procedure from $R$ random initial
strategies reduces sensitivity to the starting point. If
$\lambda_r$ is the strategy returned by restart $r$, we set
\[
  B_R=\max_{1\leq r\leq R}S_{\lambda_r}
  \leq\Lnorm[M],\qquad B_0^{\mathrm{BB}}=B_R.
\]
Here $B_0^{\mathrm{BB}}$ is the initial lower bound passed to the
exact search. A suboptimal see-saw result can affect runtime,
but cannot invalidate the subsequent BB computation.

Figure~\ref{seesaw_time} reports the quality of this initialization
for square matrices of order $n=20,25,\ldots,70$, with random
integer entries in $[-100,100]$. For each order, 50 matrices were
generated. For every matrix and every restart budget
$R=100,200,\ldots,3000$, the complete $R$-restart procedure was
run 50 times. Each run was normalized by the exact local bound
of the same matrix. Thus, at fixed $(n,R)$, the plotted minimum,
median, and arithmetic mean are computed over the pooled set of
$50\times50=2500$ ratios
\[
  \rho_{ij}(n,R)=\frac{B_R^{(i,j)}}{\Lnorm[M^{(i)}]},
  \qquad i,j=1,\ldots,50,
\]
where $i$ labels matrices and $j$ labels repeated runs. The ratio measures the attained Bell value relative to the optimum.

The data show an overall improvement with increasing restart
budget, with larger matrices generally requiring more restarts
to attain comparable ratios. At $n=70$ and $R=100$, the minimum,
median, and mean ratios are $90.96\%$, $98.22\%$, and $98.07\%$,
respectively. At $R=3000$, they are $97.02\%$, $100\%$, and
$99.75\%$. The reported median at this budget is $100\%$ for
every sampled order, although the minimum and mean ratios need not
equal $100\%$. These results support the use of see-saw
optimization to obtain a strong initial lower bound. They do
not provide a worst-case approximation guarantee. The observed
fluctuations between restart budgets also need not be monotone:
monotonicity is guaranteed only when additional restarts extend
the same retained set of candidates.

\subsection{Greedy row reordering}
\label{sec:IBB-reordering}
Greedy row reordering aims to reduce the coefficients
$c_k$, allowing more branches to be pruned. We use the fact
that permuting the rows of a matrix does not change its local
bound $\Lnorm[M]$. However, it changes which rows form the
bottom blocks used to calculate the coefficients. We therefore
rearrange the rows, starting from the bottom of the matrix.

First, we choose the row with the smallest Manhattan norm as
the last row. We then try each of the remaining $n-1$ rows
above it, calculate the local bound of each resulting two-row
matrix, and keep the row giving the smallest value. Next, we
compare the $n-2$ possible three-row matrices and again keep
the row giving the smallest local bound. We continue in this
way, adding one row at a time above the rows already selected.
When $r<n$ bottom rows have been selected, their exact local
bound gives $c_{n-r}$. The same procedure applies to rectangular
$n\times m$ matrices.

For the example in Eq.~\eqref{example_matrix}, choosing the
row with the smallest original index whenever there is a tie
gives the order $(M_3,M_2,M_4,M_1)$:
\begin{equation}
 \label{rearranged_matrix}
 \begin{gathered}
  \widetilde M=\begin{pmatrix*}[r]
   3 & -3 &  2 &  3\\
   3 &  2 & -3 & -3\\
   0 & -3 &  3 &  2\\
   2 &  3 &  3 &  0
  \end{pmatrix*},\\[.4ex]
  c=(c_1,c_2,c_3,c_4)=(21,10,8,0).
 \end{gathered}
\end{equation}
The last row has Manhattan norm $8$, while the last two and
three rows have local bounds $10$ and $21$, respectively.
Thus $c_1$ decreases from $24$ to $21$ and $c_2$ from $19$ to
$10$, while the local bound of the full matrix remains
$\Lnorm[\widetilde M]=26$. An initial strategy found before
reordering can still be used by rearranging Alice's signs in
the same way.

This procedure is greedy: at each step, it chooses the best
next row without reconsidering the rows already selected.
It therefore does not guarantee the best overall ordering or
the shortest runtime. Calculating the candidate bounds also
takes time, so it can be useful to stop after selecting a
prescribed number of bottom rows and leave the remaining rows
in their original order.

For larger bottom blocks, see-saw estimates can make the row
selection less expensive. However, these estimates are used
only to choose the ordering. The coefficients used for pruning
must still be calculated exactly by BB or replaced by valid
upper bounds. A see-saw lower estimate cannot be used directly
as a pruning coefficient. The implementation and the choice
of how many rows to reorder are discussed in
Sec.~\ref{sec:singleIBB}.

\subsection{Single-thread performance}
\label{sec:singleIBB}

\FloatBarrier
\subsubsection{Random-matrix benchmark protocol}
\label{sec:random-benchmark-protocol}

The preprocessing steps described in
Secs.~\ref{sec:IBB-seesaw} and~\ref{sec:IBB-reordering}
take extra time, but can shorten the subsequent search. To study
this balance, we vary the preprocessing depth $r$. For BB, $r$
is the number of tail coefficients computed before the search,
with the original row order retained. For IBB, it is the number
of bottom rows selected by greedy reordering, together with the
corresponding tail coefficients. Exact tail optima are precomputed only for the selected shorter
tails. At the remaining depths, any coefficient used for pruning
must still satisfy $c_k\geq\Lnorm[M_{k+1:n}]$. A coefficient
larger than the exact tail optimum is safe, but may allow fewer
branches to be discarded. Setting $c_k=+\infty$ disables pruning
at that depth.

\begin{figure*}[!tp]
  \centering
  \includegraphics[
    width=\textwidth,
    height=0.75\textheight,
    keepaspectratio
  ]{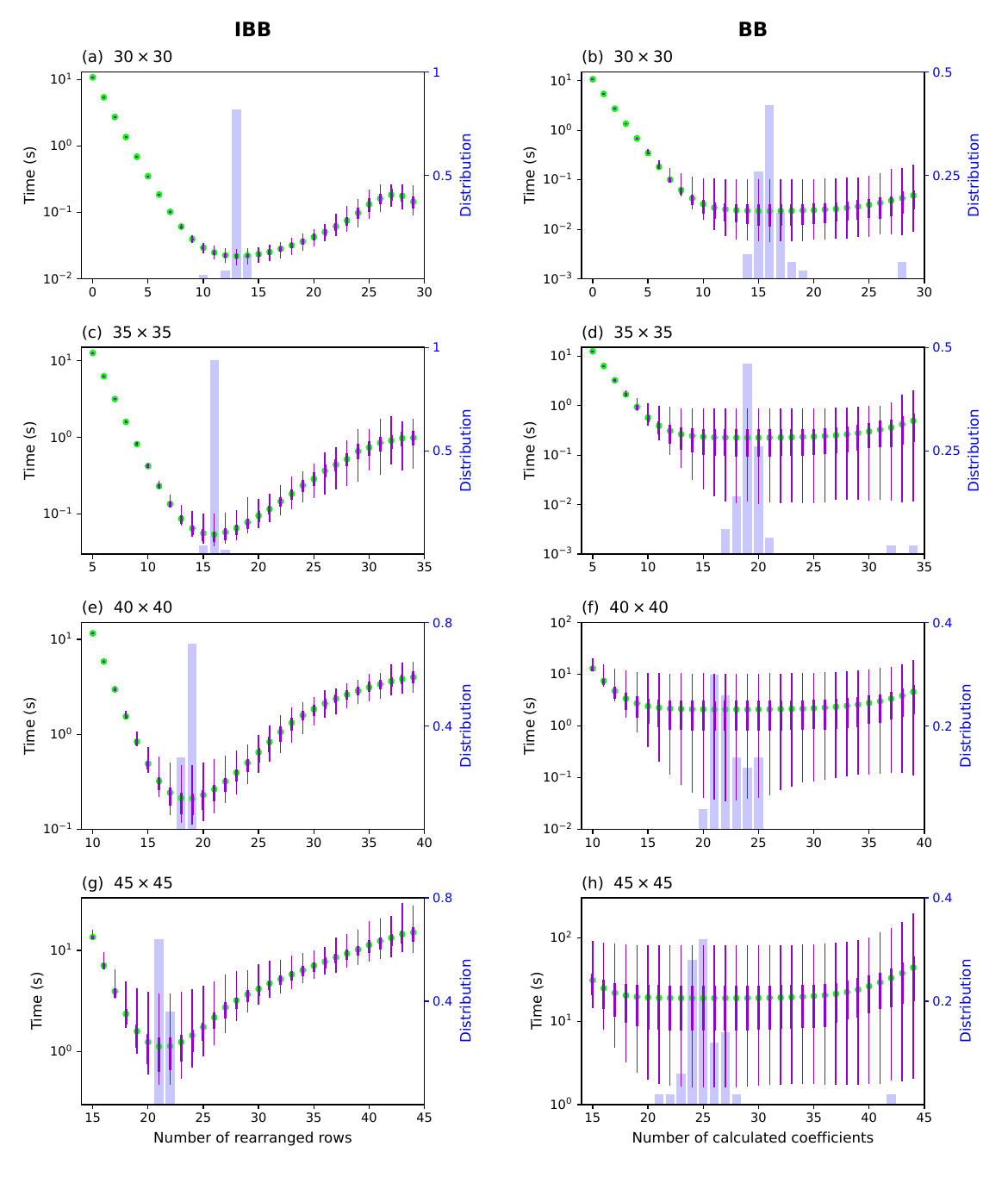}
  \caption{Single-thread runtimes for random matrices of order
  $n=30,35,40,45$, arranged from top to bottom. IBB is shown
  on the left, in panels (a), (c), (e), and (g); BB is shown
  on the right, in panels (b), (d), (f), and (h). The horizontal
  axis gives the preprocessing depth $r$. Green markers show
  mean runtimes over 50 matrices. Purple boxes span the first
  to third quartiles, and whiskers extend from the minimum to
  the maximum. Pale-blue histograms show the fraction of
  matrices whose fastest tested setting is $r$, read against
  the blue right-hand axes. Times use the logarithmic left-hand
  axes, whose limits are set separately in each panel. Each
  matrix--setting time is the shortest of three runs.}
  \label{BB_runtime}
\end{figure*}

\begin{figure*}[!tp]
  \centering
  \includegraphics[
    width=\textwidth,
    height=0.75\textheight,
    keepaspectratio
  ]{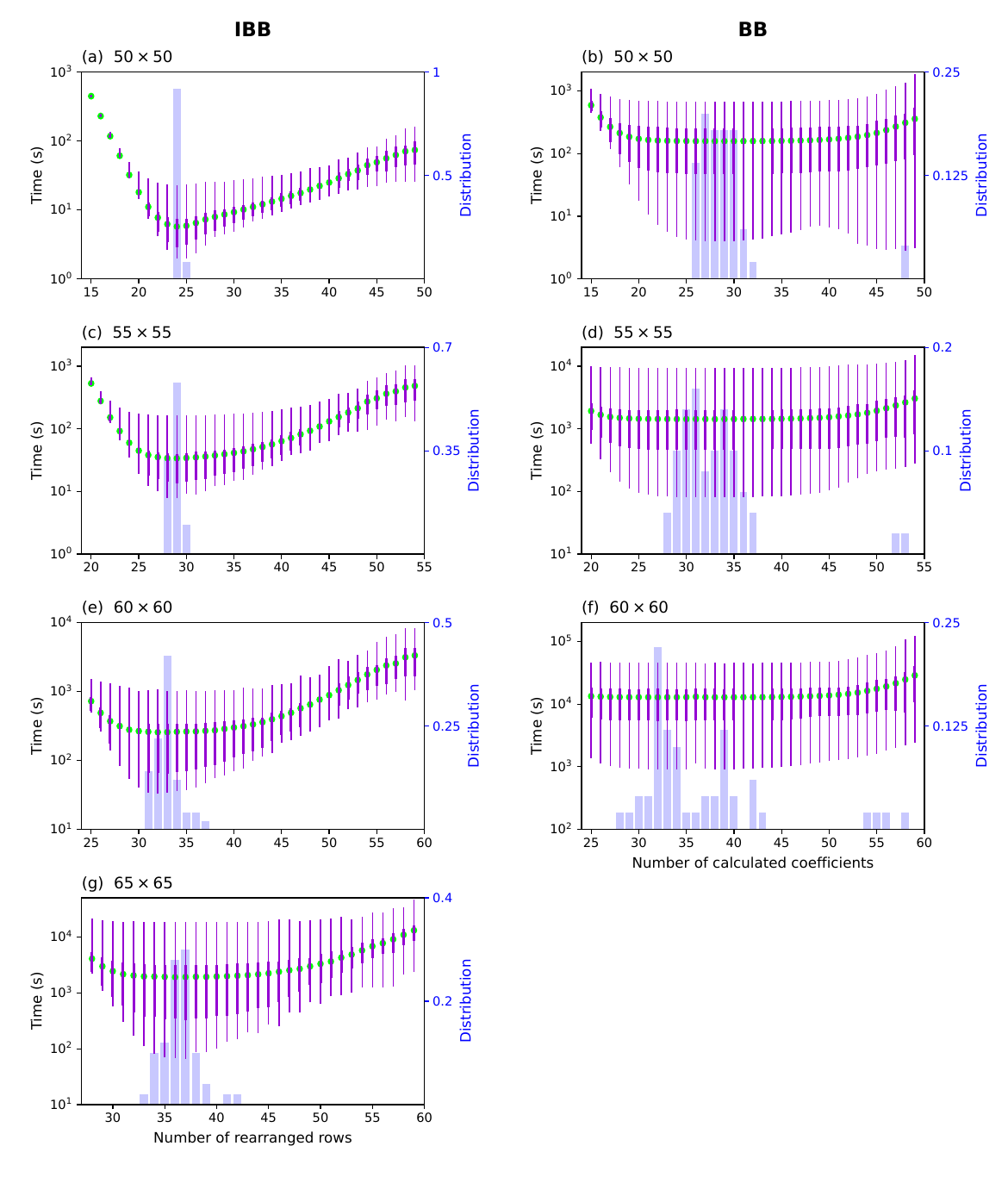}
  \caption{Single-thread runtimes for larger random matrices,
  using the same symbols and histogram convention as
  Fig.~\ref{BB_runtime}. Panels (a), (c), (e), and (g) show
  IBB for $n=50,55,60,65$, respectively; panels (b), (d), and
  (f) show BB for $n=50,55,60$. No BB measurements are reported
  for $n=65$. Each panel summarizes 50 matrices, using the
  shortest of three runs for each matrix and setting.
  Runtime axes are logarithmic and are scaled separately
  in each panel; the overlaid pale-blue histograms use the
  blue right-hand axes.} 
  \label{IBB_runtime}
\end{figure*}

We tested square $n\times n$ matrices with random integer
entries in $[-100,100]$. For each size, we generated 50 matrices
and used the same matrices for both algorithms wherever their
tested sizes overlap. Figures~\ref{BB_runtime}
and~\ref{IBB_runtime} show results for
$n=30,35,\ldots,60$ for BB and $n=30,35,\ldots,65$ for IBB.
Figure~\ref{Times} also includes the available results for
$n=20,25$. All calculations used one thread on an Intel Xeon
Gold 6448Y processor. The C code was compiled with GCC using
the \texttt{-O3} optimization option. We measured the total wall-clock runtime using the Linux
\texttt{time} command. The see-saw initialization in the IBB
benchmarks used 3000 random starting strategies.

For each matrix and each tested value of $r$, we ran the
calculation three times and kept the shortest time. Writing
$t_{A,i,\ell}(r)$ for repeat $\ell$ on matrix $i$ with algorithm
$A\in\{\mathrm{BB},\mathrm{IBB}\}$, we use
\begin{equation}
  \label{eq:runtime-statistics}
  \begin{aligned}
    t_{A,i}(r)&=\min_{\ell=1,2,3}t_{A,i,\ell}(r),\\
    \overline t_A(n,r)&=\frac{1}{50}\sum_{i=1}^{50}t_{A,i}(r).
  \end{aligned}
\end{equation}
The plots therefore show how runtimes vary between the 50
matrices after reducing the effect of fluctuations between
repeated runs. These averages should be distinguished from
the timings of individual constructed matrices discussed next.

\FloatBarrier
\subsubsection{Benchmarks with explicit constructions}
\label{sec:explicit-benchmarks}

We also tested three explicitly constructed Bell matrices.
The two $63\times63$ matrices, $G_{63}$ and $M_{63}$, are an
integer Gram matrix of the projective $E_7$ roots and a version
with modified diagonal entries, respectively. Both constructions
appear in Table~I of Ref.~\cite{DesignollePRA202602}, in the
row with $d=7$ and $m=63$. The third example is the
$72\times72$ Fishburn--Reeds matrix
$F_{72}$~\cite{FishburnSIAMJDM1994}. These examples allow a
comparison beyond random matrices, although they cover only
a small selection of constructed instances.

Table~\ref{tab:structured-timing} gives the single-thread
wall-clock times. Both algorithms used $r=35$ for the two
$63\times63$ matrices and $r=50$ for $F_{72}$. These are
timings at the stated settings, rather than averages over
matrices or the best times from a scan over $r$. IBB is about
$24$, $67$, and $128$ times faster than BB for $G_{63}$,
$M_{63}$, and $F_{72}$, respectively. The two $63\times63$
matrices take very different times despite their related
construction. Matrix size alone is therefore a poor guide
to the runtime of a particular instance.

\begin{table}[t]
  \centering
  \caption{Single-thread runtimes for three constructed Bell
  matrices at preprocessing depth $r$. Times are in seconds;
  the speedup is $t_{\mathrm{BB}}/t_{\mathrm{IBB}}$.}
  \label{tab:structured-timing}
  \begingroup
  \setlength{\tabcolsep}{2pt}
  \renewcommand{\arraystretch}{1.15}
  \begin{tabular*}{\columnwidth}
    {@{\extracolsep{\fill}}lrrrrr@{}}
    \toprule
    Matrix & Size & $r$ & BB & IBB & Speedup \\
    \midrule
    $G_{63}$ & $63\times63$ & 35
      & $2\,652$ & $111$ & $23.89$ \\
    $M_{63}$ & $63\times63$ & 35
      & $104\,055$ & $1\,548$ & $67.22$ \\
    $F_{72}$ & $72\times72$ & 50
      & $1\,545\,684$ & $12\,075$ & $128.01$ \\
    \bottomrule
  \end{tabular*}
  \endgroup
\end{table}

\subsubsection{Adaptive construction of the tail bounds}

IBB first runs the see-saw procedure on the full matrix with
3000 random starting strategies. This provides a good initial
lower bound, and the program records how long this step takes.
It then starts selecting rows from the bottom upwards, using
BB to compare the possible tails exactly. These calculations
are fast for short tails, but become more expensive as the
tails grow.

When a candidate-tail calculation takes longer than the initial
see-saw run on the full matrix, the program switches to
see-saw estimates to choose the remaining rows. After each row
is chosen, BB still computes the selected tail's coefficient
exactly. The estimates thus guide the row order but are never
used as pruning bounds. This keeps the final result exact.

The histograms in Figs.~\ref{BB_runtime} and~\ref{IBB_runtime}
help show which preprocessing settings work best. For each
matrix, we find the tested value of $r$ with the shortest
runtime and add one count to that bin. If several settings
tie, we choose the smallest $r$. Dividing the counts by 50
gives the plotted fractions, which sum to one in each panel.
The histograms show \emph{where} the best settings lie;
the purple boxes show \emph{how long} the calculations take.

\subsubsection{Dependence on preprocessing depth}

For IBB, the mean runtime usually falls rapidly as more rows
are included in preprocessing. It reaches a minimum and then
rises as the extra preprocessing costs more time than it saves
in the search. BB generally has a broader, flatter minimum.
Both trends reflect the same balance: weak tail bounds leave
more branches to explore, whereas stronger bounds take longer
to compute. More preprocessing therefore does not always
mean a faster calculation.

The best settings for the mean IBB runtime are
$r=13,16,19,21,24$ for $n=30,35,40,45,50$, respectively,
and $r=29,32,36$ for $n=55,60,65$. Thus, in these tests,
the best depth moves from slightly below half the matrix
order to slightly above it. These values apply to the tested
random matrices; other Bell matrices may require different
settings.

The best settings for individual matrices are also more tightly
grouped for IBB than for BB at the common sizes. For example,
at $n=50$, 46 of the 50 matrices have their fastest IBB run
at $r=24$, whereas the most frequent BB setting is best for
only 10 matrices. The IBB histogram becomes broader for larger
matrices. Even so, a well-defined preferred setting does not
mean that all matrices take similar times. Runtimes can still
vary considerably, and the figures show no simple rule for
how this variation changes with $r$.

\subsubsection{Optimized mean runtime and empirical growth}

\begin{figure}[t]
  \centering
  \includegraphics[width=\columnwidth]{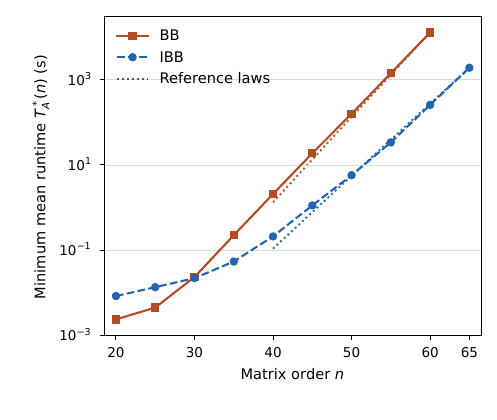}
  \caption{Best mean runtimes over the tested preprocessing
  depths, $T_A^*(n)$, for BB (squares, solid line) and IBB
  (circles, dashed line). Each mean uses 50 matrices and the
  shortest of three runs for each matrix and setting.
  Dotted curves show the reference laws in
  Eq.~\eqref{eq:runtime-reference-laws}, from $n=40$ to the
  largest tested size for each algorithm. Connecting lines
  guide the eye; the reference curves describe the measured
  trend and are not complexity bounds.}
  \label{Times}
\end{figure}

To compare BB and IBB at each matrix size, we choose the tested
setting that gives the smallest mean runtime:
\begin{equation}
  \label{eq:optimized-mean-runtime}
  T_A^*(n)=\min_{r\in\mathcal R_A(n)}\overline t_A(n,r),
\end{equation}
where $\mathcal R_A(n)$ is the set of tested depths for
algorithm $A$. We first average over the 50 matrices and then
choose $r$. This gives one setting for the whole sample,
rather than a separate best setting for each matrix.
Figure~\ref{Times} shows the resulting mean runtimes.

IBB gives a smaller best mean runtime at every common tested
size from $n=30$ to $60$. The speedup is about $1.06$ at
$n=30$, $9.88$ at $n=40$, $27.5$ at $n=50$, and $50.0$
at $n=60$. For $n=60$, the mean falls from $12\,854$~s
with BB to $257.0$~s with IBB. For IBB at $n=65$, the
best mean is $1912$~s.

At $n=20,25$, BB is faster. For these small matrices, the
search is too short for the savings to make up for IBB's
extra preprocessing, including the 3000 see-saw restarts.
This initial cost becomes less important as the matrices
grow and the exact search takes longer.

The reference curves are
\begin{equation}
  \label{eq:runtime-reference-laws}
  \begin{aligned}
    f_{\mathrm{BB}}(n)&=8.15412\times10^{-10}\,n\,1.55^n\ \mathrm{s},\\
    f_{\mathrm{IBB}}(n)&=9.52766\times10^{-10}\,n\,1.45^n\ \mathrm{s}.
  \end{aligned}
\end{equation}
They give a useful description of the growth at the larger
tested sizes, but do not closely follow all small-matrix
results. In particular, the factors $1.55$ and $1.45$ describe
these measurements; they do not establish how the algorithms
scale for arbitrarily large matrices or in the worst case.
The worst-case search remains exponential, and the gains
observed for random matrices need not carry over to a
particular constructed Bell matrix.

\subsubsection{Accounting for search progress}

Because pruning depends strongly on the matrix, elapsed time
alone gives little indication of how much work remains. A
progress indicator can instead track how many complete
strategies have already been dealt with, either by direct
evaluation or by pruning.

With $a_1=+1$ fixed, there are $2^{n-1}$ complete Alice
strategies. A node at depth $k$ has $k$ assigned signs and
contains $2^{n-k}$ possible completions. If $C$ counts the
strategies already covered, the updates are
\begin{equation}
  \label{eq:search-progress}
  \begin{aligned}
    C&\leftarrow C+1 &&\text{for an evaluated leaf},\\
    C&\leftarrow C+2^{n-k} &&\text{for a pruned subtree},\\
    P&=\frac{C}{2^{n-1}} &&\text{(fraction covered)}.
  \end{aligned}
\end{equation}
Every evaluated leaf contributes one, even if it does not
improve the best value. A leaf discarded by the pruning test
also contributes one through the second rule and is counted
only once. Since each strategy is covered once, $P$ increases
from zero to one.

This indicator measures the fraction of the strategy space
covered, rather than the fraction of the runtime completed.
One pruning test can account for a very large subtree, so
$P$ cannot be used directly to predict the remaining time.
It refers only to the final search and excludes preprocessing.

\subsubsection{Choosing the preprocessing depth}
\label{sec:optimal-reordered-rows}

We do not yet have a reliable way to predict the runtime of
an individual matrix. Such a prediction could help choose
$r$, or decide whether to work with $M$ or its transpose,
which has the same $\Lnorm$ value. As a practical starting
point, we use the random-matrix results to choose a default
preprocessing depth.

\begin{figure}[t]
  \centering
  \includegraphics[width=\columnwidth]{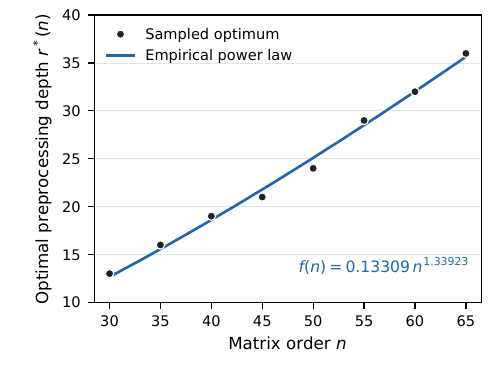}
  \caption{Preprocessing depth giving the smallest mean IBB
  runtime over the tested settings. Markers show $r^*(n)$
  for samples of 50 random matrices, with the shortest of
  three runs used for each matrix and setting. The solid
  curve is the power law in Eq.~\eqref{eq:reordering-power-law},
  shown over the tested range $30\leq n\leq65$.}
  \label{Optimal_rearranged_rows}
\end{figure}

For each order $n$, let $r^*(n)$ be the tested depth with
the smallest mean IBB runtime, choosing the smaller depth
if there is a tie:
\begin{equation}
  r^*(n)=\min\Bigl\{\operatorname*{arg\,min}_{r\in\mathcal R(n)}
  \overline t_{\mathrm{IBB}}(n,r)\Bigr\}.
  \label{eq:optimal-reordered-rows}
\end{equation}
Here $\mathcal R(n)$ is the set of tested depths. As above,
we minimize the mean over the 50 matrices, rather than average
their individual best depths. Figure~\ref{Optimal_rearranged_rows}
shows the results for $n=30,35,\ldots,65$ and the power law
\begin{equation}
  f(n)=0.13309\,n^{1.33923}.
  \label{eq:reordering-power-law}
\end{equation}
This curve approximates the observed best depths over the
tested range. It does not predict the runtime or the best
setting for every matrix.

The user can set $r$ explicitly. If no value is supplied,
the default rule is
\begin{equation}
  r_{\mathrm{def}}(n)=
  \begin{cases}
    \lfloor f(n) \rceil, & n\leq200,\\
    n-40, & n>200,
  \end{cases}
  \label{eq:default-reordered-rows}
\end{equation}
Here $n$ is the row count used by the implementation after
any matrix reduction. For $n\leq200$, the first branch rounds
the power law to the nearest integer; for example, it gives
$r_{\mathrm{def}}(120)=81$ and $r_{\mathrm{def}}(200)=161$.
The power law cannot be extended indefinitely, because $r$
cannot exceed $n$: rounding $f(400)$ would give $406$.
For $n>200$, the program instead uses the simple fallback
$r=n-40$, giving $r_{\mathrm{def}}(400)=360$.
These choices outside the tested range remain heuristics;
the benchmarks reported here extend only to $n=65$.

Choosing $r$ always involves a balance. Too little preprocessing
can leave a long search, whereas too much can spend time on
bounds that bring little further benefit. Even where exact
tail coefficients are not computed, valid upper bounds may
still allow pruning. The default value of $r$ is an empirical choice and need not give the shortest runtime for a particular matrix. If a better setting is known for a matrix or matrix family, it can be supplied explicitly.

\FloatBarrier
\subsection{Parallel IBB algorithm}
\label{sec:paraIBB}

The parallel version of IBB uses OpenMP to share the search
among several threads. It does this in two steps. First, it
divides the strategy tree into separate subtrees and assigns
them to the threads. During the search, work stealing allows
a thread that has finished its current work to take a pending
subtree from another thread.

The see-saw initialization runs serially, whereas the BB
computations used in preprocessing are parallelized with
OpenMP. All preprocessing is completed before the main
parallel search begins. All threads use the same matrix and tail coefficients, but
each keeps its own array of pending nodes. Within its assigned
subtree, each thread follows the depth-first search described
in Sec.~\ref{sec:BB-recap}.

\subsubsection{Initial partition of the strategy tree}

A sequence of initial branch choices identifies a subtree:
it contains all complete strategies that begin with that
sequence. We call this sequence a \emph{prefix}. To explain
the partition, first consider a tree with $q\geq2$ choices
at each branching step and a requested number of threads
$t\geq1$. For the local-bound calculation used here, $q=2$.
The ternary example below illustrates the same construction
for $q=3$.

If $t=q^\ell$, the partition is straightforward: assign one
of the $q^\ell$ prefixes of length $\ell$ to each thread.
Usually, however, the number of available threads is not a
power of $q$. We then start with the largest such power that
does not exceed $t$ and split some of its subtrees once more.
Splitting one subtree into $q$ children increases the number
of tasks by $q-1$. Thus we choose
\begin{equation}
  \label{num_of_branch}
  \begin{aligned}
    \ell&=\lfloor\log_q t\rfloor,\\
    s&=\left\lfloor\frac{t-q^\ell}{q-1}\right\rfloor,
  \end{aligned}
\end{equation}
where $s$ is the number of subtrees to split. The resulting
number of tasks is
\begin{equation}
  \label{num_of_new_branch}
  \begin{aligned}
    P&=q^\ell+(q-1)s\leq t,\\
    0&\leq t-P<q-1.
  \end{aligned}
\end{equation}
We split the first $s$ prefixes in their usual numerical
order, starting from the left of the tree. The assigned
prefixes then have either $\ell$ or $\ell+1$ digits. Every
complete strategy belongs to exactly one task, so no part
of the search is omitted or assigned twice.

For example, $q=3$ and $t=11$ give $\ell=2$ and $s=1$.
We begin with nine length-two prefixes and replace the first
one by its three children, giving eleven tasks
(Fig.~\ref{parallel_BB}). For $t=10$, the same rule gives
nine tasks and leaves one thread initially idle. These
partitions assume that enough branching levels remain.
If there are fewer complete strategies than threads, not
all threads can receive an initial task.

\tikzset{
  ibb tree/.style={
    x=0.118\columnwidth,y=8mm,
    every node/.style={font=\scriptsize,align=center},
    line cap=round,line join=round
  },
  ibb dot/.style={circle,fill=black,inner sep=0pt,minimum size=2.4pt},
  ibb task/.style={draw=black,rounded corners=1pt,fill=white,
    inner xsep=1.5pt,inner ysep=2pt,minimum width=5.5mm},
  ibb edge/.style={->,line width=0.55pt},
  ibb digit/.style={fill=white,inner sep=0.8pt}
}

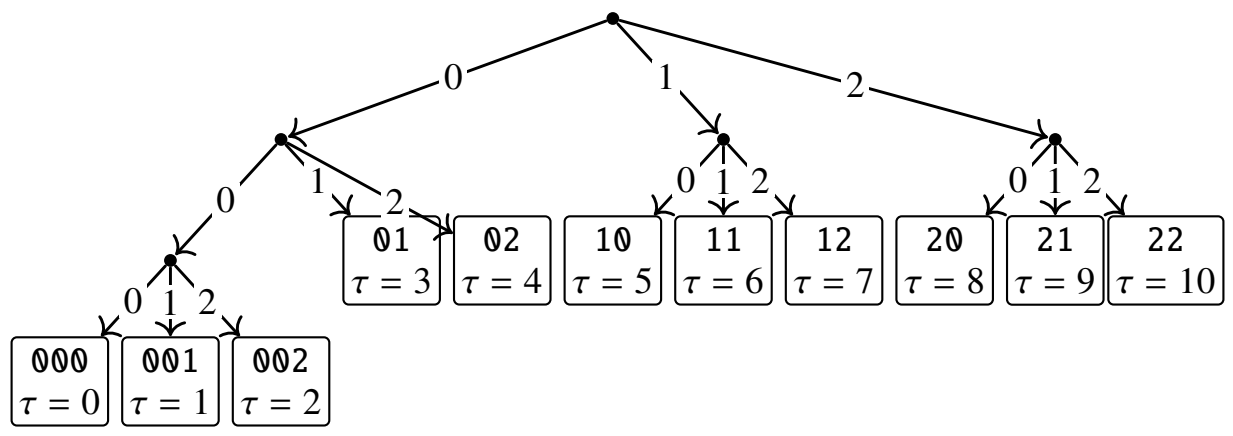
\begin{figure}[tb]
  \centering
  \begin{tikzpicture}[ibb tree]
    \path[use as bounding box] (-0.4,0.25) rectangle (7.4,-3.65);
    \node[ibb dot] (r) at (3.5,0) {};
    \node[ibb dot] (p0) at (1.4,-1) {};
    \node[ibb dot] (p1) at (4.2,-1) {};
    \node[ibb dot] (p2) at (6.3,-1) {};
    \node[ibb dot] (p00) at (0.7,-2) {};
    \node[ibb task] (p01) at (2.1,-2) {$\mathtt{01}$\\$\tau=3$};
    \node[ibb task] (p02) at (2.8,-2) {$\mathtt{02}$\\$\tau=4$};
    \node[ibb task] (p10) at (3.5,-2) {$\mathtt{10}$\\$\tau=5$};
    \node[ibb task] (p11) at (4.2,-2) {$\mathtt{11}$\\$\tau=6$};
    \node[ibb task] (p12) at (4.9,-2) {$\mathtt{12}$\\$\tau=7$};
    \node[ibb task] (p20) at (5.6,-2) {$\mathtt{20}$\\$\tau=8$};
    \node[ibb task] (p21) at (6.3,-2) {$\mathtt{21}$\\$\tau=9$};
    \node[ibb task] (p22) at (7,-2) {$\mathtt{22}$\\$\tau=10$};
    \node[ibb task] (p000) at (0,-3) {$\mathtt{000}$\\$\tau=0$};
    \node[ibb task] (p001) at (0.7,-3) {$\mathtt{001}$\\$\tau=1$};
    \node[ibb task] (p002) at (1.4,-3) {$\mathtt{002}$\\$\tau=2$};
    \draw[ibb edge] (r) -- node[ibb digit,pos=.48] {$0$} (p0);
    \draw[ibb edge] (r) -- node[ibb digit,pos=.48] {$1$} (p1);
    \draw[ibb edge] (r) -- node[ibb digit,pos=.55] {$2$} (p2);
    \draw[ibb edge] (p0) -- node[ibb digit,midway] {$0$} (p00);
    \draw[ibb edge] (p0) -- node[ibb digit,midway] {$1$} (p01);
    \draw[ibb edge] (p0) -- node[ibb digit,pos=.65] {$2$} (p02);
    \draw[ibb edge] (p1) -- node[ibb digit,midway] {$0$} (p10);
    \draw[ibb edge] (p1) -- node[ibb digit,midway] {$1$} (p11);
    \draw[ibb edge] (p1) -- node[ibb digit,midway] {$2$} (p12);
    \draw[ibb edge] (p2) -- node[ibb digit,midway] {$0$} (p20);
    \draw[ibb edge] (p2) -- node[ibb digit,midway] {$1$} (p21);
    \draw[ibb edge] (p2) -- node[ibb digit,midway] {$2$} (p22);
    \draw[ibb edge] (p00) -- node[ibb digit,midway] {$0$} (p000);
    \draw[ibb edge] (p00) -- node[ibb digit,midway] {$1$} (p001);
    \draw[ibb edge] (p00) -- node[ibb digit,midway] {$2$} (p002);
  \end{tikzpicture}
  \caption{Dividing a ternary strategy tree among eleven
  threads. Starting from nine two-digit prefixes, we split
  $\mathtt{00}$ into $\mathtt{000}$, $\mathtt{001}$, and
  $\mathtt{002}$. Each box gives the prefix of an assigned
  subtree and its thread index $\tau$, counted from zero.
  Filled dots mark nodes that are split further. The eleven
  tasks cover the complete search without overlap.}
  \label{parallel_BB}
\end{figure}

To assign a prefix to thread $\tau$, we use its base-$q$
representation. The $qs$ longer prefixes have codes
$0,\ldots,qs-1$, written with $\ell+1$ digits. The remaining
prefixes have codes $s,\ldots,q^\ell-1$, written with $\ell$
digits. Leading zeros must be kept, since they specify
branch choices. Algorithm~\ref{algorithm_distribute} gives
the assignment.

In the binary implementation, the program first limits the
number of workers to the number of available strategies.
Fixing $a_1=+1$ leaves $2^{n-1}$ complete strategies for an
$n$-row matrix, so the actual worker count is
\[
  t=\min\{t_{\mathrm{requested}},\,2^{n-1}\}.
\]
This value is used to calculate $\ell$, $s$, and $P$.
For example, a three-row matrix uses at most four workers,
even if more are requested. Each launched worker receives
an initial task, so $P=t$ and the
\texttt{NO\_INITIAL\_TASK} case does not arise. The \texttt{NO\_INITIAL\_TASK} return is retained for the
general $q$-ary case, where the partition may give $P<t$.

\begin{algorithm}[tb]
  \caption{Assigning an initial subtree to thread $\tau$}
  \label{algorithm_distribute}
  \footnotesize
  \begin{algorithmic}[1]
    \Require $q\geq2$, $t\geq1$, and thread index $0\leq\tau<t$
    \Statex Use $\ell,s,P$ from
      Eqs.~\eqref{num_of_branch}--\eqref{num_of_new_branch}.
    \If{$\tau\geq P$}
     \State \Return \texttt{NO\_INITIAL\_TASK}
      \EndIf
    \If{$\tau<qs$}
      \State $J\gets\tau$, \quad $d\gets\ell+1$
    \Else
      \State $J\gets\tau-(q-1)s$, \quad $d\gets\ell$
    \EndIf
    \If{$d=0$}
      \State \Return the empty prefix
    \EndIf
    \State Write $J$ in base $q$ using exactly $d$ digits,
    \Statex \hspace{\algorithmicindent}including leading zeros.
    \State \Return the resulting prefix $(u_1,\ldots,u_d)$
  \end{algorithmic}
\end{algorithm}

For $\Lnorm[M]$, the tree is binary. Fixing $a_1=+1$
leaves $n-1$ signs to choose, and the rule above assigns one
initial task to every thread whenever $t\leq2^{n-1}$.
A binary digit $0$ means sign $-1$, and $1$ means sign $+1$.
For a prefix $(u_1,\ldots,u_d)$, the initial node is therefore
\begin{equation}
  \label{eq:parallel-node-initialization}
  \begin{aligned}
    S&=(+1,2u_1-1,\ldots,2u_d-1),\\
    k&=d+1,\qquad v=\sum_{x=1}^{k}a_xM_x.
  \end{aligned}
\end{equation}

\begin{figure}[tb]
  \centering
  \begin{minipage}{\columnwidth}
    \textbf{(a) Five threads}\par\smallskip
    \centering
    \begin{tikzpicture}[ibb tree]
      \path[use as bounding box] (-0.4,0.25) rectangle (7.4,-3.65);
      \node[ibb dot] (r) at (3.5,0) {};
      \node[ibb dot] (p0) at (1.5,-1) {};
      \node[ibb dot] (p1) at (5.5,-1) {};
      \node[ibb dot] (p00) at (0.5,-2) {};
      \node[ibb task] (p01) at (2.5,-2) {$\mathtt{01}$\\$\tau=2$};
      \node[ibb task] (p10) at (4.5,-2) {$\mathtt{10}$\\$\tau=3$};
      \node[ibb task] (p11) at (6.5,-2) {$\mathtt{11}$\\$\tau=4$};
      \node[ibb task] (p000) at (0,-3) {$\mathtt{000}$\\$\tau=0$};
      \node[ibb task] (p001) at (1,-3) {$\mathtt{001}$\\$\tau=1$};
      \draw[ibb edge,blue] (r) -- node[ibb digit,midway] {$0$} (p0);
      \draw[ibb edge,orange] (r) -- node[ibb digit,midway] {$1$} (p1);
      \draw[ibb edge,blue] (p0) -- node[ibb digit,midway] {$0$} (p00);
      \draw[ibb edge,orange] (p0) -- node[ibb digit,midway] {$1$} (p01);
      \draw[ibb edge,blue] (p1) -- node[ibb digit,midway] {$0$} (p10);
      \draw[ibb edge,orange] (p1) -- node[ibb digit,midway] {$1$} (p11);
      \draw[ibb edge,blue] (p00) -- node[ibb digit,midway] {$0$} (p000);
      \draw[ibb edge,orange] (p00) -- node[ibb digit,midway] {$1$} (p001);
    \end{tikzpicture}
  \end{minipage}
  \par\medskip
  \begin{minipage}{\columnwidth}
    \textbf{(b) Six threads}\par\smallskip
    \centering
    \begin{tikzpicture}[ibb tree]
      \path[use as bounding box] (-0.4,0.25) rectangle (7.4,-3.65);
      \node[ibb dot] (r) at (3.5,0) {};
      \node[ibb dot] (p0) at (1.5,-1) {};
      \node[ibb dot] (p1) at (5.5,-1) {};
      \node[ibb dot] (p00) at (0.5,-2) {};
      \node[ibb dot] (p01) at (2.5,-2) {};
      \node[ibb task] (p10) at (4.5,-2) {$\mathtt{10}$\\$\tau=4$};
      \node[ibb task] (p11) at (6.5,-2) {$\mathtt{11}$\\$\tau=5$};
      \node[ibb task] (p000) at (0,-3) {$\mathtt{000}$\\$\tau=0$};
      \node[ibb task] (p001) at (1,-3) {$\mathtt{001}$\\$\tau=1$};
      \node[ibb task] (p010) at (2,-3) {$\mathtt{010}$\\$\tau=2$};
      \node[ibb task] (p011) at (3,-3) {$\mathtt{011}$\\$\tau=3$};
      \draw[ibb edge,blue] (r) -- node[ibb digit,midway] {$0$} (p0);
      \draw[ibb edge,orange] (r) -- node[ibb digit,midway] {$1$} (p1);
      \draw[ibb edge,blue] (p0) -- node[ibb digit,midway] {$0$} (p00);
      \draw[ibb edge,orange] (p0) -- node[ibb digit,midway] {$1$} (p01);
      \draw[ibb edge,blue] (p1) -- node[ibb digit,midway] {$0$} (p10);
      \draw[ibb edge,orange] (p1) -- node[ibb digit,midway] {$1$} (p11);
      \draw[ibb edge,blue] (p00) -- node[ibb digit,midway] {$0$} (p000);
      \draw[ibb edge,orange] (p00) -- node[ibb digit,midway] {$1$} (p001);
      \draw[ibb edge,blue] (p01) -- node[ibb digit,midway] {$0$} (p010);
      \draw[ibb edge,orange] (p01) -- node[ibb digit,midway] {$1$} (p011);
    \end{tikzpicture}
  \end{minipage}
  \caption{Dividing the binary search among (a) five and
  (b) six threads. Each box gives an assigned prefix and
  its thread index $\tau$. The four two-digit prefixes are
  first formed, then the leftmost one or two are split
  further. Blue edges marked $0$ mean sign $-1$; orange
  edges marked $1$ mean sign $+1$. These choices follow the
  fixed first sign $a_1=+1$. The tasks cover the whole search
  without overlap.}
  \label{parallel_BB_binary}
\end{figure}
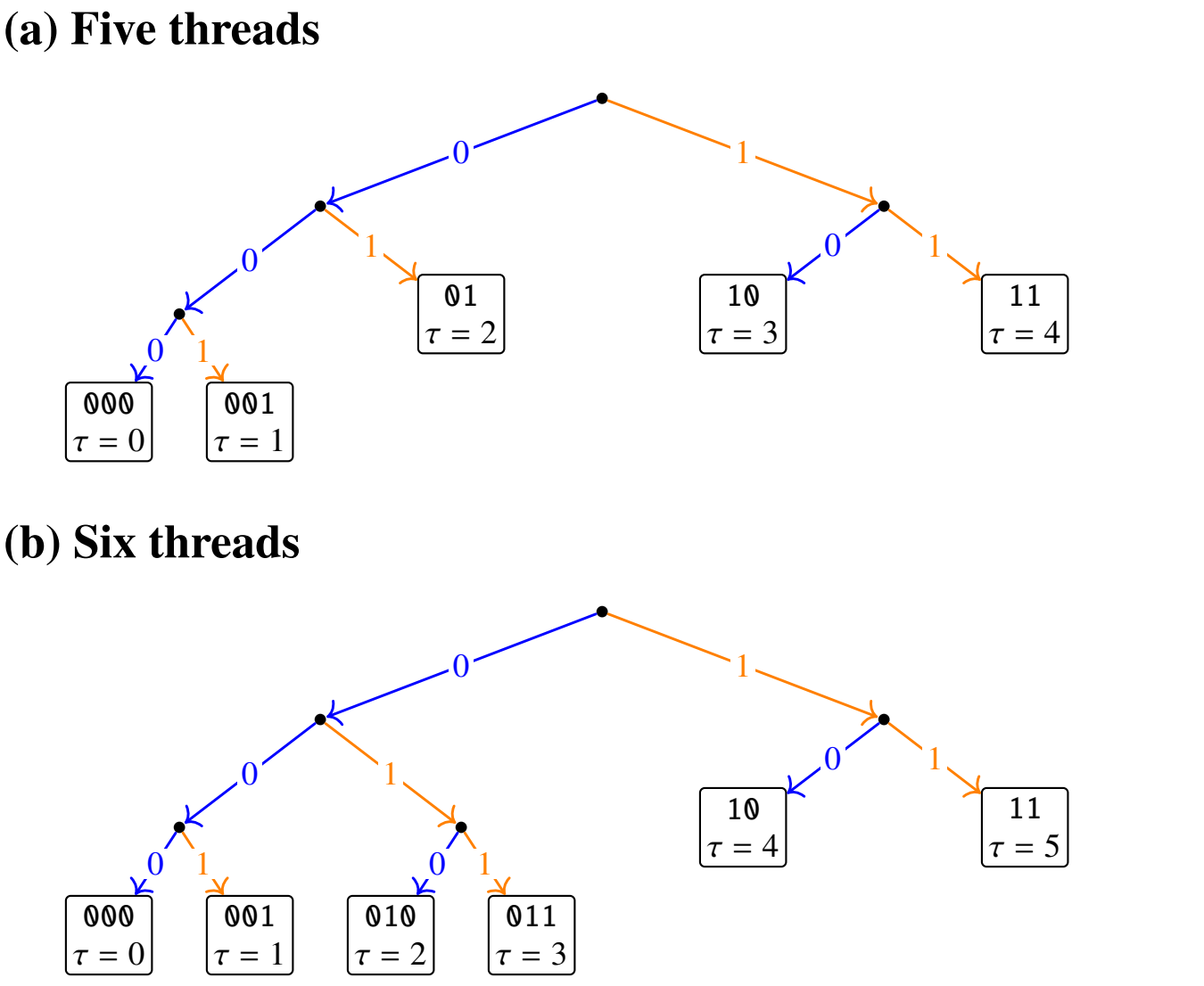

Figure~\ref{parallel_BB_binary} shows the assignments for
five and six threads. With five threads, for example,
$\ell=2$ and $s=1$. Threads 0 and 1 receive the three-digit
prefixes $\mathtt{000}$ and $\mathtt{001}$; threads 2, 3,
and 4 receive $\mathtt{01}$, $\mathtt{10}$, and $\mathtt{11}$.
For thread 0, the reordered matrix in
Eq.~\eqref{rearranged_matrix} gives
$S=(+1,-1,-1,-1)$, $k=4$, and $v=(-2,-5,-1,4)$.
Before pruning, tasks of these two prefix lengths differ
in size by a factor of two. Their actual runtimes can differ
much more, because some subtrees are easier to prune than others.

\subsubsection{Dynamic load balancing}

The initial partition does not ensure that all threads finish
at the same time. A thread may quickly prune most of its
subtree while another still has a long search ahead. Work
stealing lets a thread that becomes idle take over some of
this remaining work.

Each thread stores its pending nodes in a private array.
Two indices mark the part of the array still in use:
$b$ is the first occupied position and $h$ is the last.
They correspond to \texttt{counter\_back} and
\texttt{counter} in the code. As in Sec.~\ref{sec:memory},
the thread normally processes the last record, at $h$.
Its array is empty when $b>h$.

After a prescribed number of search operations, a busy thread
checks whether another thread is idle. If it can pass on
work, it copies the record at $b$ into a shared transfer
buffer and increases $b$ by one. The receiving thread copies
that record into its own array and sets $b=h=0$. The transfer
is synchronized so that the donated subtree has only one
owner and is neither lost nor processed twice.

The donated record is taken from the beginning of the occupied
array, rather than from the end where the busy thread is
working. These older pending nodes are usually closer to
the root and contain larger subtrees. Giving one of them
to an idle thread can therefore keep it busy longer and
reduce the number of transfers. Transfers themselves take
time, and frequent access to the shared buffer can become
a bottleneck. The interval between checks must therefore
balance quick redistribution against this extra cost.

Each thread prunes using a valid lower bound $B$, such as
the value of a known complete strategy. Sharing a better
value can help other threads prune more branches, but a
thread can safely continue with an older, smaller bound.
The final answer is the largest value obtained by any thread.
The search is finished only when all threads are idle and
no work remains in transit.

Table~\ref{work_steal} shows a transfer for the original
matrix in Eq.~\eqref{example_matrix}. With records written
as $(k,v,S)$, consider the two pending nodes
\[
  \begin{aligned}
    R_-&=(2,(-1,1,6,3),(+1,-1)),\\
    R_+&=(2,(5,5,0,-3),(+1,+1)).
  \end{aligned}
\]
Thread 1 keeps $R_+$ and passes $R_-$ to thread 2. During
this transfer, only thread 1's lower index $b$ increases;
its upper index $h$ stays unchanged. No records need to be
shifted within the array.

\begin{table}[tb]
  \centering
  \caption{Passing a pending subtree from thread 1 to an
  idle thread 2. Entries show the pending records and the
  occupied array interval $[b,h]$. In the middle row, $R_-$
  is in the shared transfer buffer. Thread 2 then copies it
  into its own array. The buffer is empty before and after
  the transfer, and the best known value can remain $B=24$.}
  \label{work_steal}
  \begingroup
  \footnotesize
  \setlength{\tabcolsep}{3pt}
  \renewcommand{\arraystretch}{1.2}
  \begin{tabular*}{\columnwidth}{@{\extracolsep{\fill}}lcc@{}}
    \toprule
    Stage & Thread 1 & Thread 2\\
    \midrule
    Before & $R_-,R_+\quad[0,1]$ & empty $[0,-1]$\\
    Publish $R_-$ & $R_+\quad[1,1]$ & empty $[0,-1]$\\
    Receive $R_-$ & $R_+\quad[1,1]$ & $R_-\quad[0,0]$\\
    \bottomrule
  \end{tabular*}
  \endgroup
\end{table}

\FloatBarrier
\section{Applications}
\label{sec:results}
The exact evaluation of classical correlation bounds is a
central bottleneck in constructing Bell inequalities with a
large quantum advantage. We apply IBB to three families of
explicit Bell matrices. The first yields improved lower bounds
on finite-order Grothendieck constants for $d=3,4,5,7$.
The second consists of two Fishburn--Reeds matrices, for which
IBB confirms the conjectured classical bounds for $k=9,10$.
The third consists of matrices with coefficients restricted
to $\{\pm1\}$, defining uniform-input XOR games whose
quantum-to-classical bias ratios exceed the CHSH value
$\sqrt2$. The largest square instance solved here has size $130\times130$;
the rectangular instances include a $72\times360$ matrix. These sizes
describe the computations reported here, rather than a general size
limit: the search effort depends strongly on the matrix structure.

\subsection{Lower bounds on finite-order Grothendieck constants}
\label{sec:applications-grothendieck}
For $M\ne0$, let $\Lnorm[M]$ and $Q(M,d)$ be the classical and
vector bounds defined in Eqs.~\eqref{CB} and~\eqref{eq:quantum-vector-bound}.
The order $d$ denotes the real vector dimension. 
By Eq.~\eqref{eq:kg-lower}, every feasible vector strategy
with value $q_d(M)$ yields a lower bound on $K_G(d)$:
\begin{equation}
 \label{eq:applications-ratio}
 \rho_d(M):=\frac{q_d(M)}{\Lnorm[M]}
 \leq\frac{Q(M,d)}{\Lnorm[M]}\leq K_G(d).
\end{equation}
Thus a globally optimal quantum strategy is not required for a
Grothendieck lower bound. IBB determines the binary optimum in the
denominator; the strategy values in the numerator are evaluated separately. Throughout this section, decimal strategy values and ratios are numerical approximations, whereas the reported integer classical bounds are exact IBB results. For all Bell matrices evaluated below with IBB, Table~\ref{Grothendieck_table} lists the classical bounds $\Lnorm[M]$, the resulting lower bounds on $K_G(d)$, and the wall-clock times required to compute $\Lnorm[M]$.

\begin{table*}[t]
 \centering
 \caption{Explicit Bell matrices evaluated with IBB. A single
 subscript denotes a square matrix of that order; two indices
 specify its numbers of rows and columns. The table lists the
 vector dimension $d$, preprocessing depth $r$, exact classical bound $\Lnorm[M]$,
 numerical strategy value $q_d$, ratio
 $\rho_d=q_d/\Lnorm[M]$, and wall-clock time.
 Strategy values and ratios are approximate; boldface marks
 improvements over the cited reference bounds. Timings refer
 to 64-core runs on an HP Z8 G5 workstation.
 A dash indicates an unreported timing.}
 \label{tab:matrix_comparison}
 \label{Grothendieck_table}
 \begingroup
 \small
 \setlength{\tabcolsep}{5pt}
 \renewcommand{\arraystretch}{1.15}
 \begin{tabular*}{\textwidth}
   {@{\extracolsep{\fill}}lrrrrrr@{}}
 \toprule
 Matrix & $d$ & $r$ & $\Lnorm[M]$ & $q_d$ & $\rho_d$ & Time\\
 \midrule
 $M_{97}$ & 3 & 61 & $1\,392\,129$
   & $2.000072\times10^6$
   & 1.436700 & $0{:}27{:}56$\\[2pt]
 $M_{81\times321}$ & 3 & 48 & $320\,975\,385$
   & $4.618022\times10^8$
   & \textbf{1.438746} & $0{:}02{:}38$\\[2pt]
 $M_{72\times360}$ & 4 & 41 & $164\,850\,797$
   & $2.449834\times10^8$
   & \textbf{1.486092} & $2{:}20{:}29$\\[2pt]
 $M_{80\times269}$ & 5 & 47 & $919\,782\,264$
   & $1.378665771\times10^9$
   & \textbf{1.498904} & $110{:}00{:}45$\\[2pt]  
 $M_{91}$ & 7 & 56 & $524\,482$
   & $7.865485\times10^5$
   & \textbf{1.499667} & $13{:}22{:}20$\\[2pt]
 $F_{90}$ & 10 & 55 & $3060$
   & $4500$ & 1.470588 & $15{:}56{:}53$\\
 \bottomrule
 \end{tabular*}
 \par\smallskip
 \parbox{\textwidth}{
   Times are given as $\mathrm{h}{:}\mathrm{min}{:}\mathrm{s}$.
 }
 \endgroup
\end{table*}

\paragraph{Fishburn--Reeds matrices}
The family of matrices $F_{k(k-1)}$ is obtained from the projective $D_k$ root
configuration with a diagonal modification~\cite{FishburnSIAMJDM1994}.
For this family, the ratio of the $k$-dimensional vector-strategy
value to the classical bound is conjectured to be $(3k-5)/(2k-3)$. Fishburn and Reeds established the optimal
diagonal modification for $k=3,4,5$; Designolle \emph{et al.}
extended the exact computations to
$k=6,7,8$~\cite{DesignollePRA202602}. Our IBB calculations determine the exact classical bounds $\Lnorm[F_{72}]$ and $\Lnorm[F_{90}]$. We list them alongside the separately obtained vector-strategy values:
\begin{equation}
 \label{eq:applications-fr}
 \begin{aligned}
  \Lnorm[F_{72}]&=2160, & q_9(F_{72})&=3168,\\
  \Lnorm[F_{90}]&=3060, & q_{10}(F_{90})&=4500.
 \end{aligned}
\end{equation}
The resulting ratios are $22/15$ and $25/17$, respectively, in
agreement with the conjectured expression for $k=9,10$.
These instances test the solver on structured matrices, however, they do
not improve the cited Grothendieck lower bounds at these orders.
The $F_{72}$ instance also appears in the single-thread benchmark
comparison.

\paragraph{Square instances in dimensions 3 and 7}
The $97\times97$ instance $M_{97}$ is an integer approximation to
the Bell matrix constructed using Frank--Wolfe methods in
Ref.~\cite{Designolle2023} and subsequently studied in
Ref.~\cite{DesignollePRA202602}. The earlier work solved its
high-precision representative using a QUBO solver; here IBB gives
$\Lnorm[M_{97}]=1\,392\,129$ for the integer instance used in this
work. Its ratio reproduces the reference value $1.43670$ to the
precision shown. The $91\times91$ instance $M_{91}$ starts from
the Gram matrix of the union of the $63$ projective $E_7$ roots
and a real equiangular tight frame of $28$ lines in $\mathbb R^7$.
A Frank--Wolfe separating hyperplane, followed by rescaling and
integer rounding, produces the Bell matrix; it is not obtained
solely by modifying the Gram matrix's diagonal. Its binary optimum
was treated heuristically in Ref.~\cite{DesignollePRA202602}.
The exact value $\Lnorm[M_{91}]=524\,482$ now supports a ratio
of approximately $1.499667$. The underlying constructions are
available in the accompanying dataset~\cite{DesignolleData2024}.

\paragraph{Rectangular instances}
The three rectangular instances considered here were supplied
by S.~Designolle. Their construction follows the Frank--Wolfe
approach of Refs.~\cite{Designolle2023,DesignollePRA202602}, implemented in
\texttt{BellPolytopes.jl}~\cite{BellPolytopesJL}. Starting from
a scalar-product matrix of two sets of unit vectors, this
procedure finds a separating Bell inequality for the local
correlation polytope. Rescaling and integer rounding then
produce the integer Bell matrices evaluated with IBB.

The matrix $M_{72\times360}$ uses projective configurations
in $\mathbb{R}^4$. Alice's $72$ directions comprise the $60$
lines of the $600$-cell and the $12$ lines of a dual $24$-cell.
Bob's $360$ directions comprise the same $60$ lines and the
$300$ lines of the $120$-cell. This extends the
$60\times360$ construction of
Ref.~\cite{DesignollePRA202602}. IBB gives the exact classical
bound $\Lnorm[M_{72\times360}]=164\,850\,797$.

The matrix $M_{81\times321}$ uses two nested spherical grids
in $\mathbb{R}^3$. Starting from an icosahedron, each refinement
adds the normalized midpoint $(x+y)/\|x+y\|$ of every edge
of the current triangulation. The successive grids contain
$12$, $42$, $162$, and $642$ vertices, corresponding to
$6$, $21$, $81$, and $321$ directions after identifying
antipodal points. Alice uses the $81$ directions at refinement
level two, and Bob uses the $321$ directions at level three.
IBB gives the exact classical bound
$\Lnorm[M_{81\times321}]=320\,975\,385$.

The third instance, $M_{80\times269}$, targets $d=5$.
Its exact classical bound is
$\Lnorm[M_{80\times269}]=919\,782\,264$, obtained with IBB
in approximately $4.6$ days on 64 cores. Together with the
numerical strategy value
$q_5\simeq1.3786657706\times10^9$, this gives
$\rho_5\simeq1.498904$.

The corresponding matrix files are named
\texttt{M72x360}, \texttt{M81x321}, and \texttt{M80x269},
respectively.

Table~\ref{Grothendieck_table} summarizes the results.
The ratios for $M_{81\times321}$, $M_{72\times360}$,
$M_{80\times269}$, and $M_{91}$ improve on the reference
lower bounds $1.43670$, $1.48579$, $1.49339$, and $1.49339$
for $d=3,4,5,7$, respectively, reported in
Ref.~\cite{DesignollePRA202602}. For $d=7$, the reference
value follows from monotonicity, $\KG(7)\geq\KG(5)$.
The displayed decimals are rounded numerical ratios.

\subsection{Signed Bell games beyond the CHSH bias ratio}
\label{app:signed_games}
A signed matrix $M\in\{\pm1\}^{n\times n}$ defines a uniform-input
XOR game: the referee samples $(i,j)$ uniformly, and the players
win when the product of their outputs equals $M_{ij}$~\cite{Cleve2004}.
Its optimal classical bias, defined as twice the winning probability
minus one, is $\beta_{\mathrm C}=\Lnorm[M]/n^2$. A vector strategy
of value $q_d(M)$ has bias $\beta_q=q_d(M)/n^2$, so that
\begin{equation}
 \label{eq:applications-game-bias}
 \frac{\beta_q}{\beta_{\mathrm C}}=\rho_d(M).
\end{equation}
The comparison below concerns bias ratios, not ratios of winning probabilities. For the CHSH matrix
\begin{equation}
 M_{\mathrm{CHSH}}=\begin{pmatrix}1&1\\1&-1\end{pmatrix},
\end{equation}
the quantum-to-classical ratio is
\begin{equation}
 \frac{Q(M_{\mathrm{CHSH}},2)}{\Lnorm[M_{\mathrm{CHSH}}]}
 =\frac{2\sqrt2}{2}=\sqrt2.
\end{equation}
We note that random signed games have also been studied in this framework in Ref.~\cite{Ambainis2012}.

Restricting coefficients to $\{\pm1\}$ does not change the
supremum defining $K_G(d)$, or $K_G$ in the unrestricted-dimensional
limit. This follows from a block approximation and the random
rounding estimate of Lov\'asz~\cite[Lemma~10.11 and Eq.~(10.9)]{Lovasz2012}:
a real matrix can be replaced by a sufficiently large signed matrix
with arbitrarily small error in the relevant normalized values.
The computational question is therefore to find comparatively
small explicit examples whose classical bounds can be evaluated
exactly. We give five such examples, of orders $100$, $108$, $120$,
$120$, and $130$, with four-dimensional bias ratios above~$\sqrt2$.

\paragraph{Geometric constructions}
Four instances use a deformed super-Fibonacci distribution of unit
vectors on $S^3$, inspired by Ref.~\cite{Alexa2022}. For
$i=1,\ldots,n$, set
\begin{equation}
 \begin{aligned}
 t_i&=\frac{i-1+\delta}{n-1+2\delta}, &
 r_i&=\frac{t_i^p}{t_i^p+(1-t_i)^p},\\
 \theta_i&=\frac{2\pi(i-1)}{\alpha}, &
 \psi_i&=\frac{2\pi(i-1)}{\beta}.
 \end{aligned}
\end{equation}
Then define
\begin{equation}
 \label{eq:super_fibonacci_vectors}
 v_i=\begin{pmatrix}
 \sqrt{r_i}\sin\theta_i\\
 \sqrt{r_i}\cos\theta_i\\
 \sqrt{1-r_i}\sin\psi_i\\
 \sqrt{1-r_i}\cos\psi_i
 \end{pmatrix}\in\mathbb R^4.
\end{equation}
For each vector set, form its Gram matrix
$G_{ij}=\langle v_i,v_j\rangle$ and apply the signed cubic kernel
\begin{equation}
 \label{eq:signed_cubic_kernel}
 M_{ij}=\operatorname{sgn}_{+}(G_{ij}-cG_{ij}^3),
 \qquad
 \operatorname{sgn}_{+}(z)=
 \begin{cases}+1,&z\geq0,\\-1,&z<0.\end{cases}
\end{equation}
Note that cubic kernels are motivated by recent numerical investigations of large quantum-to-classical ratios~\cite{Bene2026}. The parameters in Table~\ref{tab:super-fibonacci-parameters}
produce $M_{100}$, $M_{108}$, $M_{120}$, and $M_{130}$. In particular, the last row lists the values of the parameter $c$
used in the signed cubic kernel of Eq.~\eqref{eq:signed_cubic_kernel}. The fifth instance, $C_{120}$ uses all $120$ vertices of the regular $600$-cell, normalized to unit length. Coordinates are given in Table~1 of Ref.~\cite{Waegell2010}; include both members of each antipodal pair and divide the listed coordinates by two. For $C_{120}$, we use $c=1.275$ and assign exact zero inner products the sign $+1$. 

\begin{table}[tb]
 \centering
 \caption{Parameters of the four deformed super-Fibonacci
 constructions. The displayed values define the vectors and the signed cubic kernel; $\alpha=\sqrt2$ for $M_{108}$ is used exactly.}
 \label{tab:super-fibonacci-parameters}
 \begingroup
 \fontsize{7.5}{9.2}\selectfont
 \setlength{\tabcolsep}{2pt}
 \renewcommand{\arraystretch}{1.12}
 \begin{tabular*}{\columnwidth}{@{\extracolsep{\fill}}lrrrr@{}}
 \toprule
 Parameter & $M_{100}$ & $M_{108}$ & $M_{120}$ & $M_{130}$\\
 \midrule
 $\alpha$ & 1.4147222445 & $\sqrt2$ & 1.4161756990 & 1.3879796802\\
 $\beta$ & 1.5349312127 & 1.5337511688 & 1.5297081964 & 1.5452062615\\
 $\delta$ & 0.6632991799 & $1/2$ & 0.6649311431 & 0.4484168767\\
 $p$ & 0.9840827988 & 1 & 1.0947604461 & 1.0220494380\\
 $c$ & 1.2660079049 & 1.2856774795 & 1.2850408574 & 1.2673929097\\
 \bottomrule
 \end{tabular*}
 \endgroup
\end{table}

\paragraph{Quantum strategies and exact classical bounds}
The generating vectors give a feasible value
\begin{equation}
 q_4(M)=\sum_{i,j=1}^{n}M_{ij}\langle v_i,v_j\rangle
 \leq Q(M,4).
\end{equation}
Further numerical optimization may allow Alice's and Bob's vectors to vary independently in $\mathbb R^4$. The ratios $\rho_4$ reported in Table~\ref{tab:signed_games} exceed those obtained directly from the generating vectors for all four super-Fibonacci instances. For $C_{120}$, on the other hand, the generating vectors reproduce the listed value to numerical precision. This agreement alone does not establish global quantum optimality, and none is needed for the ratios reported here.

\begin{table}[t]
 \centering
 \caption{Signed Bell games with four-dimensional bias ratios
 exceeding $\sqrt2$. Every matrix is square, with order given by
 its subscript, and has only $\pm1$ entries. $r$ denotes preprocessing depth. The integer classical
 bounds are exact IBB results; the ratios $\rho_4=q_4/\Lnorm[M]$
 are numerical approximations. Wall-clock times refer
 to the classical-bound computation on 64 cores and are written
 as $\mathrm{h}{:}\mathrm{min}{:}\mathrm{s}$.}
 \label{tab:signed_games}
 \begingroup
 \small
 \setlength{\tabcolsep}{5pt}
 \renewcommand{\arraystretch}{1.15}
 \begin{tabular*}{\columnwidth}{@{\extracolsep{\fill}}lrrrr@{}}
 \toprule
 Matrix & $r$ & $\Lnorm[M]$ & $\rho_4\ \text{(approx.)}$ & Time\\
 \midrule
 $M_{100}$ & 63 & 2512 & 1.426270010 & $0{:}01{:}43$\\
 $M_{108}$ & 70 & 2860 & 1.418321312 & $0{:}04{:}13$\\
 $M_{120}$ & 81 & 3448 & 1.436887152 & $0{:}18{:}01$\\
 $C_{120}$ & 81 & 3712 & 1.449336715 & $2{:}09{:}32$\\
 $M_{130}$ & 90 & 4016 & 1.445477824 & $0{:}39{:}51$\\
 \bottomrule
 \end{tabular*}
 \endgroup
\end{table}

All five examples exceed the CHSH bias ratio. The smallest,
$M_{100}$, has ratio approximately $1.42627$; the largest ratio
among these examples is approximately $1.44934$, attained by the
$600$-cell strategy for $C_{120}$. These explicit instances show
that coefficients of equal magnitude suffice to surpass the CHSH
bias ratio at matrix sizes accessible to exact IBB evaluation.

Finally, the four-dimensional vector strategies have a direct
quantum realization~\cite{TsirelsonJoSM1987}. Let
$\Gamma_1,\ldots,\Gamma_4$ be traceless Hermitian $4\times4$
matrices satisfying
$\Gamma_k\Gamma_l+\Gamma_l\Gamma_k=2\delta_{kl}I_4$.
On the maximally entangled state
\begin{equation}
 |\Phi_4\rangle=\frac12\sum_{k=1}^{4}|k,k\rangle,
\end{equation}
choose the dichotomic observables
\begin{equation}
 A_i=\sum_{k=1}^{4}(u_i)_k\Gamma_k,
 \qquad
 B_j=\sum_{k=1}^{4}(v_j)_k\Gamma_k^{\mathsf T}.
\end{equation}
They satisfy
$\langle\Phi_4|A_i\otimes B_j|\Phi_4\rangle=\langle u_i,v_j\rangle$.
Hence each reported strategy can be implemented with local
Hilbert-space dimension four, however, no claim of minimality is required.

\FloatBarrier
\section{Conclusion and outlook}
\label{sec:disc}
We have combined see-saw initialization, greedy row reordering, and OpenMP parallelism in an improved branch-and-bound algorithm for computing exact classical bounds of bipartite correlation Bell inequalities. In our approach, heuristics guide the search, while rigorous upper bounds ensure that pruning preserves exactness.

Our benchmarks highlight the importance of balancing preprocessing against search effort. For the random $60\times60$ ensemble, we reduced the optimized mean runtime from approximately $12\,854$~s with BB to $257$~s with IBB, a speedup of about 50. For the representative single-thread structured benchmarks, we obtained speedups of about 24, 67, and 128 for $G_{63}$, $M_{63}$, and $F_{72}$, respectively. Our tests also show the limits of these improvements: the gains depend on the matrix, and preprocessing can cost more time than it saves for small instances. The worst-case search remains exponential.

The exact local bounds support improved Grothendieck lower bounds for
$d=3,4,5,7$, with the numerical ratios reported in
Table~\ref{Grothendieck_table}. In particular, the rectangular instances $M_{81\times321}$, $M_{72\times360}$, and $M_{80\times269}$ yield ratios of approximately $1.438746$, $1.486092$, and $1.498904$ for $d=3$, $d=4$, and $d=5$, respectively, while $M_{91}$ yields approximately $1.499667$ for $d=7$. For the
Fishburn--Reeds family, the exact classical bounds at matrix sizes 72 and 90
agree with the conjectured ratios $22/15$ and $25/17$, extending the
cases evaluated in Ref.~\cite{DesignollePRA202602} to $k=9,10$.
Five explicit signed matrices additionally give four-dimensional
quantum-to-classical bias ratios above $\sqrt2$. The largest ratio
among these examples is approximately $1.449337$, obtained from the
$600$-cell construction, and the largest square instance solved has
order 130. Each reported four-dimensional vector strategy has a
realization on a maximally entangled state of two ququarts.

Several questions remain open. Choosing the preprocessing depth automatically, using simple properties of the matrix or information collected early in the search, could avoid repeated runs with different parameter settings. The search might also be shortened by changing the row order, using the transposed matrix, finding tighter upper bounds for the remaining rows, or exploiting matrix symmetries. Measuring how the runtime changes with the number of CPU cores would help quantify the time spent redistributing work and sharing the best value found. Finally, compact records of the search and the bounds used to discard branches would make the results easier to check independently. Carefully bounding numerical errors in the quantum-strategy values would provide an additional level of verification.

A concrete open problem is to determine the exact classical bound
$\Lnorm[U_{120}]$ of the $120\times120$ Bell matrix based on the
$E_8$ root system introduced by Designolle
et al.~\cite{DesignollePRA202602} in Table~I of that paper.
A copy of this matrix is available as \texttt{U120.txt} in our
repository. It is associated with the conjectured lower bound
$\KG(8)\geq165/109\approx1.51376$. We estimate that an exact IBB
computation would require several centuries on our available
workstation, placing it beyond our current computational resources.
An exact evaluation would determine whether this matrix certifies
the conjectured lower bound above.

Our method could also be extended to Bell expressions containing marginal terms, prepare-and-measure witnesses such as those considered in Ref.~\cite{MartonCPC2025}, and scenarios with more outcomes~\cite{Arajo2020,Falerio2026} or a limited amount of classical communication~\cite{Sidajaya2023,Marton2024,Chen2026}. 

Each extension would need branching rules and rigorous upper bounds adapted to the problem. On the physics side, a concrete goal is to find a smaller matrix with entries $\pm1$ whose classical bound can be computed exactly and for which $q_3(M)/\Lnorm[M]>\sqrt2$. The corresponding three-dimensional vector strategy would then surpass the CHSH bias ratio using a maximally entangled pair of qubits. The signed-matrix approximation argument shows that such examples exist in principle. The remaining challenge is to find small examples whose classical bounds can be computed in a reasonable time, and to determine how few measurement settings are needed.

\section*{Data availability statement}
The data and numerical scripts supporting the findings of this study are openly available at
\url{https://github.com/istvanmarton/IBB}.

\section*{Acknowledgements}
We thank S\'ebastien Designolle for fruitful discussions and for kindly providing us the large rectangular matrix instances. We thank Gábor Drótos for fruitful discussions and valuable advice. We thank M\'aty\'as Koniorczyk for fruitful discussions and for testing some of our large matrix instances using an exact QUBO solver.
I.M. was supported by the János Bolyai Research Scholarship of the Hungarian Academy of Sciences.
We acknowledge support from the European Union (CHIST-ERA MoDIC) and from the National Research, Development and Innovation Office NKFIH (Grant Nos.~2023-1.2.1-ERA\_NET-2023-00009 and K145927). The use of LLM tools was limited to improving the language, readability, and organization of the text.

\appendix
\renewcommand{\theequation}{A.\arabic{equation}}
\renewcommand{\thefigure}{A.\arabic{figure}}
\renewcommand{\thetable}{A.\arabic{table}}

\section{Branch-and-bound: a worked example}
\label{BB_algorithm_with_example}

This appendix illustrates the depth-first branch-and-bound (BB) search
through one complete example. We use the notation and the $+1$-first
branching order of the main text. The numbered nodes in
Fig.~\ref{fig:app-bb-traversal} correspond to the nine steps in
Table~\ref{tab:app-bb-trace}. Color emphasizes the three possible actions:
\textcolor{BBbranch}{\emph{branch}},
\textcolor{BBprune}{\emph{prune}}, and
\textcolor{BBupdate}{\emph{update the incumbent}}; the labels also
identify these actions in grayscale.

\subsection{The matrix, bounds, and search rule}
\label{sec:app-bb-setup}
Let $M_x$ denote row $x$ of the matrix in Eq.~\eqref{example_matrix},
reproduced here for convenience:
\begin{equation}
 M=\begin{pmatrix*}[r]
 2&3&3&0\\3&2&-3&-3\\3&-3&2&3\\0&-3&3&2
 \end{pmatrix*}.
 \label{eq:app-bb-matrix}
\end{equation}
For this four-row matrix, the local-bound formula in
Eq.~\eqref{eq:l1man} becomes
\begin{equation}
 L_1(M)=\max_{a_1,\ldots,a_4=\pm1}
 \left\lVert\sum_{x=1}^4a_xM_x\right\rVert_1,
 \label{eq:app-bb-objective}
\end{equation}
where $\|v\|_1=\sum_y|v_y|$.

Fixing $a_1=+1$ removes the
global-sign redundancy and leaves eight complete strategies.
At depth $k$, the assigned signs form a prefix
$S=(a_1,\ldots,a_k)$ and determine
$v=\sum_{x=1}^ka_xM_x$. We use the exact tail bounds
\begin{equation}
 \begin{aligned}
 c_1&=L_1(M_{2:4})=24,& c_2&=L_1(M_{3:4})=19,\\
 c_3&=\|M_4\|_1=8,& c_4&=0.
 \end{aligned}
 \label{eq:app-bb-tails}
\end{equation}
Here $M_{k+1:4}$ contains the rows \emph{after} the first $k$
assigned signs. The triangle inequality gives the subtree bound
\begin{equation}
 \left\lVert v+\sum_{x=k+1}^4a_xM_x\right\rVert_1
 \leq\underbrace{\|v\|_1+c_k}_{U_k(v)}.
 \label{eq:app-bb-upper}
\end{equation}
Thus $U_k(v)$ bounds every completion of the current prefix.

Initialize the incumbent lower bound at $B_0=c_1=24$.
This is valid because adding a row cannot decrease $L_1$:
for any signed sum $t$ of the last three rows,
\[
 \begin{aligned}
 &\max\{\|t+M_1\|_1,\|t-M_1\|_1\}\\
 &\quad\geq\tfrac12\bigl(\|t+M_1\|_1+\|t-M_1\|_1\bigr)
 \geq\|t\|_1.
 \end{aligned}
\]
Global-sign reversal puts either choice into the convention
$a_1=+1$. Maximizing over $t$ proves $L_1(M)\geq24$.
Hence 24 is a \emph{certified initial lower bound}, not necessarily
a value already obtained by evaluating a complete strategy.
This initialization uses the \emph{exact} tail optimum; an arbitrary
upper estimate of that optimum would not provide a valid lower bound.

At each visited node, test $U_k(v)\leq B$. If true, return $B$:
no completion can improve it. Otherwise, update $B=\|v\|_1$ at a
leaf ($k=4$), or branch at an internal node. In the latter case,
\begin{equation}
 \begin{aligned}
 B_+&=g_{M,c}(k+1,v+M_{k+1},B),\\
 g_{M,c}(k,v,B)&=g_{M,c}(k+1,v-M_{k+1},B_+).
 \end{aligned}
 \label{eq:app-bb-recursion}
\end{equation}
The entire $+1$ subtree is processed before the $-1$ subtree,
which receives the updated bound $B_+$. The notation
$g_{M,c}(k,v,B)$ always has three arguments.

\begin{figure*}[!t]
 \centering
 \resizebox{\textwidth}{!}{
\begin{tikzpicture}[x=1cm,y=1cm,>=Stealth,
 every node/.style={font=\small,align=center},
 box/.style={draw,rounded corners=2pt,inner sep=4pt,minimum width=2.65cm,minimum height=1.12cm},
 branch/.style={box,draw=BBbranch,fill=BBbranch!5},
 prune/.style={box,draw=BBprune,fill=BBprune!5},
 update/.style={box,draw=BBupdate,fill=BBupdate!8,line width=1pt},
 skip/.style={draw=black!30,dashed,rounded corners=1pt,text=black!55,inner sep=3pt,font=\footnotesize},
 edge/.style={->,line width=.65pt},
 lab/.style={fill=white,inner sep=1pt,font=\footnotesize},
 badge/.style={circle,fill=black,text=white,font=\bfseries\footnotesize,inner sep=2pt,minimum size=5mm}]
 \node[branch] (n1) at (0,0) {$S=(+)$\\$U=32>B=24$\\\textcolor{BBbranch}{branch}};
 \node[branch] (n2) at (-4.4,-1.95) {$S=(+,+)$\\$U=32>B=24$\\\textcolor{BBbranch}{branch}};
 \node[branch] (n7) at (4.4,-1.95) {$S=(+,-)$\\$U=30>B=26$\\\textcolor{BBbranch}{branch}};
 \node[prune] (n3) at (-6.6,-3.95) {$S=(+,+,+)$\\$U=20\leq B=24$\\\textcolor{BBprune}{prune}};
 \node[branch] (n4) at (-2.2,-3.95) {$S=(+,+,-)$\\$U=26>B=24$\\\textcolor{BBbranch}{branch}};
 \node[prune] (n8) at (2.2,-3.95) {$S=(+,-,+)$\\$U=26=B$\\\textcolor{BBprune}{prune at equality}};
 \node[prune] (n9) at (6.6,-3.95) {$S=(+,-,-)$\\$U=20\leq B=26$\\\textcolor{BBprune}{prune}};
 \node[prune,minimum width=1.9cm,font=\footnotesize] (n5) at (-3.45,-6.0) {$S=(+,+,-,+)$\\$U=12,\ B=24$\\\textcolor{BBprune}{discard leaf}};
 \node[update,minimum width=1.9cm,font=\footnotesize] (n6) at (-.95,-6.0) {$S=(+,+,-,-)$\\$U=26,\ B=24$\\\textcolor{BBupdate}{update $B\to26$}};
 \foreach \p/\l/\r in {n1/n2/n7,n2/n3/n4,n7/n8/n9,n4/n5/n6}{
  \draw[edge] (\p) -- node[lab,pos=.48] {$+1$} (\l);
  \draw[edge] (\p) -- node[lab,pos=.48] {$-1$} (\r);
 }
 \foreach \p/\x/\txt [count=\i] in {n3/-7.7/{(+,+,+,+)},n3/-5.5/{(+,+,+,-)},n8/1.1/{(+,-,+,+)},n8/3.3/{(+,-,+,-)},n9/5.5/{(+,-,-,+)},n9/7.7/{(+,-,-,-)}}{
  \node[skip] (skip\i) at (\x,-6.0) {$\txt$\\not visited};
  \draw[dashed,black!30] (\p) -- (skip\i);
 }
 \foreach \n in {1,2,3,4,5,6,7,8,9}{\node[badge,anchor=center] at (n\n.north west) {\n};}
 \node[font=\footnotesize,anchor=west,align=left] at (-8.55,-7.05)
 {\textbf{Processing order:} $1\to2\to3\to4\to5\to6\to7\to8\to9$.
 \quad Left child: $+1$ first; right child: $-1$ second.};
\end{tikzpicture}}
 \caption{Depth-first traversal for Eq.~\eqref{eq:app-bb-matrix}, with
 $a_1=+1$ fixed. Black badges give the order in which nodes are processed;
 $+$ and $-$ denote $+1$ and $-1$. Each solid box shows the prefix $S$,
 its bound $U=U_k(v)$, and the incumbent on entry. Gray dashed descendants
 are skipped, not evaluated. The update at node 6 is propagated to node 7
 and makes equality pruning possible at node 8. Consecutive visit numbers
 need not be adjacent in the tree: after a subtree ends, the search
 backtracks to the nearest pending sibling.}
 \label{fig:app-bb-traversal}
\end{figure*}

\subsection{The nine node visits}
\label{sec:app-bb-visits}
\emph{Visits 1--2: initial branching.}
At the root, $k=1$, $v=M_1=(2,3,3,0)$, and $U_1=8+24=32>24$.
The search therefore enters the $a_2=+1$ child first, obtaining
$v=(5,5,0,-3)$. Its bound is again $U_2=13+19=32>24$, so it branches
on $a_3$. The pending $a_2=-1$ child is left for later.

\emph{Visit 3: pruning an entire subtree.}
For $S=(+,+,+)$, $v=(8,2,2,0)$ and $U_3=12+8=20\leq24$.
Neither choice of $a_4$ can improve $B$. Both descendant leaves
are skipped without forming their signed row sums.

\emph{Visits 4--6: finding a better incumbent.}
Backtracking to $S=(+,+,-)$ gives $v=(2,8,-2,-6)$ and
$U_3=18+8=26>24$, so both signs of $a_4$ must be considered.
The $+1$ child, $v=(2,5,1,-4)$, has value 12 and is discarded.
The $-1$ child gives
\begin{equation}
 v=(2,11,-5,-8),\qquad \|v\|_1=26,
 \label{eq:app-bb-optimum-leaf}
\end{equation}
and updates $B$ from 24 to 26. At a leaf, $c_4=0$, so the bound
is the objective value itself.

\emph{Visits 7--9: using the stronger bound.}
The search returns to the pending prefix $S=(+,-)$, with
$v=(-1,1,6,3)$. Its bound $U_2=11+19=30>26$ still requires
branching. The $+1$ child has $v=(2,-2,8,6)$ and
$U_3=18+8=26=B$, so it is pruned. The $-1$ child has
$v=(-4,4,4,0)$ and $U_3=12+8=20\leq26$, so it is also pruned.
No pending nodes remain.

\begin{table*}[t]
 \centering
 \caption{Complete trace of the search in Fig.~\ref{fig:app-bb-traversal}.
 $B_{\rm in}$ and $B_{\rm out}$ are the incumbent immediately before and
 after processing the indicated node, excluding subsequent processing of
 its children. Thus a branching row does not change $B$ itself.}
 \label{tab:app-bb-trace}
 \small\setlength{\tabcolsep}{5pt}\renewcommand{\arraystretch}{1.16}
 \begin{tabular*}{\textwidth}{@{\extracolsep{\fill}}rcllrrrrrl@{}}
 \toprule
 Visit & $k$ & $S$ & $v$ & $\|v\|_1$ & $c_k$ & $U_k(v)$ & $B_{\rm in}$ & $B_{\rm out}$ & Action\\
 \midrule
 1&1&$(+)$&$(2,3,3,0)$&8&24&32&24&24&\textcolor{BBbranch}{Branch}\\
 2&2&$(+,+)$&$(5,5,0,-3)$&13&19&32&24&24&\textcolor{BBbranch}{Branch}\\
 3&3&$(+,+,+)$&$(8,2,2,0)$&12&8&20&24&24&\textcolor{BBprune}{Prune two leaves}\\
 4&3&$(+,+,-)$&$(2,8,-2,-6)$&18&8&26&24&24&\textcolor{BBbranch}{Branch}\\
 5&4&$(+,+,-,+)$&$(2,5,1,-4)$&12&0&12&24&24&\textcolor{BBprune}{Discard leaf}\\
 6&4&$(+,+,-,-)$&$(2,11,-5,-8)$&26&0&26&24&\textbf{26}&\textcolor{BBupdate}{Update incumbent}\\
 7&2&$(+,-)$&$(-1,1,6,3)$&11&19&30&26&26&\textcolor{BBbranch}{Branch}\\
 8&3&$(+,-,+)$&$(2,-2,8,6)$&18&8&26&26&26&\textcolor{BBprune}{Prune two leaves}\\
 9&3&$(+,-,-)$&$(-4,4,4,0)$&12&8&20&26&26&\textcolor{BBprune}{Prune two leaves}\\
 \bottomrule
 \end{tabular*}
\end{table*}

\subsection{Optimality, ties, and the pending-node stack}
\label{sec:app-bb-lessons}
The leaf at visit 6 proves $L_1(M)\geq26$. Every other strategy
is either evaluated or lies below a pruned node whose upper bound
is at most 26. Consequently,
\begin{equation}
 \boxed{L_1(M)=g_{M,c}(1,M_1,24)=26.}
 \label{eq:app-bb-result}
\end{equation}
One maximizing strategy pair is
$a=(1,1,-1,-1)$ and $b=(1,1,-1,-1)$, where $b_y=\operatorname{sgn}(v_y)$
for the vector in Eq.~\eqref{eq:app-bb-optimum-leaf}.

\emph{Equality pruning preserves the optimum, not every optimizer.}
At visit 8, the old bound $B=24$ would not allow pruning because
$U_3=26>24$. The improved incumbent makes the test succeed.
Indeed, the skipped strategy $(1,-1,1,1)$ also attains 26.
This does not affect the computed optimum. To enumerate all maximizing
strategies, one would instead prune only when $U_k(v)<B$ and retain
all tied maximizing leaves.

The main search processes nine nodes of the 15-node tree and evaluates
two of its eight leaves; three internal prunings discard the other six
leaves. These counts exclude the preprocessing used to compute $c_k$.
They illustrate pruning rather than a general runtime guarantee.

Finally, this traversal matches the stack implementation in the main
text. When an internal node branches, replace its record by the
$-1$ child and append the $+1$ child. Processing the last record then
produces the order shown in Fig.~\ref{fig:app-bb-traversal}.
After visit 3 the next active node is 4; after visit 6 it is 7,
which uses the updated incumbent 26. The stack is empty after visit 9.

\FloatBarrier
\bibliographystyle{elsarticle-num}
\bibliography{IBB_cit}

\end{document}